\documentclass{aastex702}

\newcommand{\etal}{et~al.}
\newcommand{\cgsflux}{erg~s$^{-1}$~cm$^{-2}$}

\newcommand{\kms}{\hbox{km~s$^{-1}$}}

\newcommand{\ArII}{\hbox{{\rm Ar}\kern 0.1em{\sc ii}}}
\newcommand{\ArIII}{\hbox{{\rm Ar}\kern 0.1em{\sc iii}}}
\newcommand{\CIV}{\hbox{{\rm C}\kern 0.1em{\sc iv}}}
\newcommand{\HI}{\hbox{{\rm H}\kern 0.1em{\sc i}}}
\newcommand{\HII}{\hbox{{\rm H}\kern 0.1em{\sc ii}}}
\newcommand{\HeI}{\hbox{{\rm He}\kern 0.1em{\sc i}}}
\newcommand{\HeII}{\hbox{{\rm He}\kern 0.1em{\sc ii}}}
\newcommand{\NII}{\hbox{{\rm N}\kern 0.1em{\sc ii}}}
\newcommand{\OI}{\hbox{{\rm O}\kern 0.1em{\sc i}}}
\newcommand{\OII}{\hbox{{\rm O}\kern 0.1em{\sc ii}}}
\newcommand{\OIII}{\hbox{{\rm O}\kern 0.1em{\sc iii}}}
\newcommand{\OIIlong}{{\rm O}\kern 0.1em{\sc ii}~$\lambda 3727$} 
\newcommand{\FeII}{\hbox{{\rm Fe}\kern 0.1em{\sc ii}}}
\newcommand{\NeII}{\hbox{{\rm Ne}\kern 0.1em{\sc ii}}}
\newcommand{\NeIII}{\hbox{{\rm Ne}\kern 0.1em{\sc iii}}}
\newcommand{\NeV}{\hbox{{\rm Ne}\kern 0.1em{\sc v}}}
\newcommand{\SII}{\hbox{{\rm S}\kern 0.1em{\sc ii}}}
\newcommand{\SIII}{\hbox{{\rm S}\kern 0.1em{\sc iii}}}
\newcommand{\SIV}{\hbox{{\rm S}\kern 0.1em{\sc iv}}}
\newcommand{\SiIV}{\hbox{{\rm Si}\kern 0.1em{\sc iv}}}
\newcommand{\MgII}{\hbox{{\rm Mg}\kern 0.1em{\sc ii}}}
\newcommand{\Halpha}{\hbox{{\rm H}\kern 0.1em$\alpha$}}
\newcommand{\Hbeta}{\hbox{{\rm H}\kern 0.1em$\beta$}}
\newcommand{\Heopta}{\hbox{{\rm He}\kern 0.1em{\sc i}}~$6678$}
\newcommand{\Heoptb}{\hbox{{\rm He}\kern 0.1em{\sc i}}~$5876$}
\newcommand{\Heoptc}{\hbox{{\rm He}\kern 0.1em{\sc i}}~$4471$}
\newcommand{\Brgam}{\hbox{{\rm Br}\kern 0.1em$\gamma$}}
\newcommand{\Brten}{\hbox{{\rm Br}\kern 0.1em$10$}}
\newcommand{\Breleven}{\hbox{{\rm Br}\kern 0.1em$11$}}
\newcommand{\HeIh}{\hbox{{\rm He}\kern 0.1em{\sc i}}~$1.7$~{\micron}}
\newcommand{\HeIk}{\hbox{{\rm He}\kern 0.1em{\sc i}}~$2.06$~{\micron}}
\newcommand{\squishlist}{
   \begin{list}{$\bullet$}
    { \setlength{\itemsep}{0pt}      \setlength{\parsep}{1pt}
      \setlength{\topsep}{3pt}       \setlength{\partopsep}{0pt}
      \setlength{\leftmargin}{1.5em} \setlength{\labelwidth}{1em}
      \setlength{\labelsep}{0.5em} } }
\newcommand{\squishend}{
    \end{list}  }  
\newcommand{\MJysrmicron}{MJy sr$^{-1}\mu$m}
\newcommand{\SBcgsunits}{erg s$^{-1}$ cm$^{-2}$ sr$^{-1}$}
\newcommand{\MJysr}{MJy sr$^{-1}$}
\shorttitle{Interstellar He with JWST}
\shortauthors{Rigby et al.}
\submitjournal{ApJ}
\journalinfo{Submitted to ApJ, 5 August 2026}  
\received{5 August 2026}
\begin{document}

\title{Ubiquitous Interstellar Neutral Helium Detected with JWST}

\correspondingauthor{Jane R. Rigby}

\author[orcid=0000-0002-7627-6551]{Jane R. Rigby}
\affiliation{Astrophysics Science Division, NASA Goddard Space Flight Center, Code 660, 8800 Greenbelt Rd, Greenbelt, MD 20771, USA}
\email[show]{Jane.Rigby@NASA.gov}

\author[orcid=0000-0002-5716-3412]{D. Koutroumpa}
\affiliation{LATMOS/IPSL, UVSQ Université Paris-Saclay, Sorbonne Université, CNRS, Guyancourt, 78280, France}
\email{Dimitra.Koutroumpa@latmos.ipsl.fr}

\author[orcid=0000-0003-3702-7592]{M. Galeazzi}
\affiliation{Department of Physics, University of Miami, Coral Gables, FL 33146, USA}
\email{galeazzi@miami.edu}

\author[orcid=0000-0001-6251-4988]{T. Hutchison}
\affiliation{Astrophysics Science Division, NASA Goddard Space Flight Center, Code 660, 8800 Greenbelt Rd, Greenbelt, MD 20771, USA}
\email{taylor.hutchison@nasa.gov}

\author[orcid=0000-0001-6654-5378]{K. D. Kuntz}
\affiliation{Johns Hopkins University, Department of Physics \& Astronomy, Baltimore, MD 21218, USA}
\affiliation{Astrophysics Science Division, NASA Goddard Space Flight Center, Code 660, 8800 Greenbelt Rd, Greenbelt, MD 20771, USA}
\email{kip.d.kuntz@nasa.gov}

\author[orcid=0000-0003-3351-0878]{Rosalia O'Brien}
\affiliation{Department of Astronomy, University of Maryland, College Park, MD 20742, USA}
\affiliation{Astrophysics Science Division, Code 660, NASA Goddard Space Flight Center, 8800 Greenbelt Rd., Greenbelt, MD 20771, USA}
\affiliation{Center for Research and Exploration in Space Science and Technology, NASA/GSFC, Greenbelt, MD 20771 USA}
\email{rosalia.d.obrien@nasa.gov}

\author[orcid=0000-0002-3191-8151]{Marshall Perrin}
\affiliation{Space Telescope Science Institute, 3700 San Martin Dr, Baltimore, MD 21218, USA}
\email{mperrin@stsci.edu}

\author[orcid=0000-0002-6374-1119]{F. S. Porter}
\affiliation{Astrophysics Science Division, NASA Goddard Space Flight Center, Code 660, 8800 Greenbelt Rd, Greenbelt, MD 20771, USA}
\email{frederick.s.porter@nasa.gov}

\author[orcid=0000-0003-0083-9554]{Yu. Ralchenko}
\affiliation{Astrophysics Science Division, NASA Goddard Space Flight Center, Code 660, 8800 Greenbelt Rd, Greenbelt, MD 20771, USA}
\affiliation{Department of Astronomy, University of Maryland, College Park, MD 20742, USA}
\affiliation{Center for Research and Exploration in Space Science and Technology, NASA Goddard Space Flight Center, Greenbelt, MD 20771, USA}
\email{yuri.ralchenko@nasa.gov}

\author[orcid=0000-0001-7426-5413]{B. M. Walsh}
\affiliation{Center for Space Physics, Boston University, Boston, MA, 02215, USA}
\email{bwalsh@bu.edu}

\author[orcid=0000-0003-1815-0114]{Brian Welch}
\affiliation{International Space Science Institute, Hallerstrasse 6, 3012 Bern, Switzerland}
\email{bwelch.astro@gmail.com}

\begin{abstract}
  We report the discovery of ubiquitous neutral helium emission in sky spectra taken with JWST's NIRSpec instrument. The emission, with a wavelength of 1.0833~\micron, resembles one of the ``sky lines'' that are seen by ground-based observatories. We examine this emission in all suitable NIRSpec spectra in the public archive, totaling 22~d of exposure time. We find that this He I emission is almost always present: it is well-detected  in $80\%$ of the observations, and in $53\%$ of the individual exposures.  The emission is highly time-variable: at a given pointing, the line intensity can vary by factors of several over the course of a day.  The He I emission is strongest when JWST crosses through the cone of interstellar neutral helium that is gravitationally focused by the sun;  intensity during the cone crossing is anti-correlated with solar activity.  The low redshift and narrow velocity width of the He I line, and the elevated intensity when JWST crosses through the focusing cone, together indicate that origin of the He I emission is cold Milky Way gas passing through our solar system.   JWST provides a new way to study this interstellar gas, revealing new insights such as extreme variability on timescales of hours to days, which has not been previously reported. 
\end{abstract}

\section{Introduction}

JWST \citep{Gardner.2006, Gardner.2023} is a large space telescope in orbit around the second sun-Earth Lagrange point (L2). Its NIRSpec instrument \citep{Jakobsen.2022, Boker.2023} was designed to take extremely deep spectra of galaxies in the early universe, and exceeds the sensitivity of previous observatories by one to three orders over its 0.6--5.3~\micron\ range \citep{Rigby.2023}.

\begin{figure}
\centering
\includegraphics[width=\textwidth]{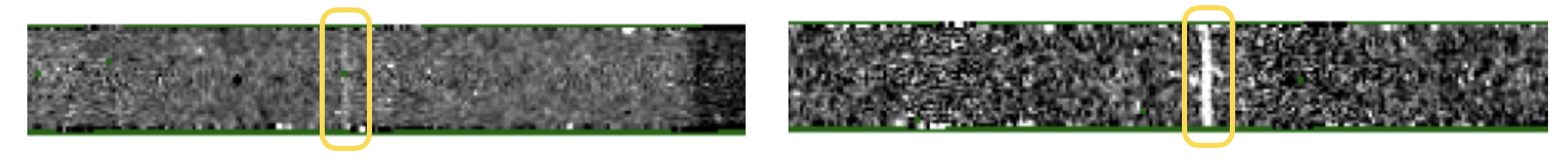}
\caption{Examples of He I 1.0833~\micron\ emission (marked with yellow boxes) in two fairly deep JWST NIRSpec exposures.  What is shown are reduced exposure--level (level 2) two-dimensional spectra (from S2D files), with wavelength increasing along the horizontal axis.  The emission fills the length of the slit, indicating that the emission is diffuse rather than from discrete sources.
Left panel: a spectrum taken with the G140M grating, F070LP blocking filter, and S200A1 fixed slit of width 200 mas wide, with effective exposure time of 2772~s, from program 03215, observation 1, visit 3.   Right panel: a spectrum taken with G140H, F100LP, S200A1 fixed slit, 3239~s, from program 03788, observation 9, visit 1.
}
\label{fig:skyline}
\end{figure}

We serendipitously discovered that NIRSpec spectra generically show faint emission from the sky (rather than from discrete sources) at a wavelength of 1.0833~micron. In two-dimensional spectra, the appearance of the feature resembles one the sky lines that contaminate spectra obtained from ground-based observatories; see Figure~\ref{fig:skyline}.
The wavelength of this emission corresponds to the 1.0833~\micron\
1s2s $^3$S – 1s2p $^3$P triplet of neutral helium\footnote{Of the dozen elements that have transitions with vacuum wavelengths between 1.083 and 1.084~\micron, helium is by far the most abundant.} (He I) \citep{kramida2024nistasd}.

He~I~1.0833~\micron\ line emission has been detected in the direction of the sun by experiments specially designed for total solar eclipses \citep{Kuhn.1996, Judge.2019, Molnar.2025}, as well as in polarized light through coronagraphy \citep{Kuhn.2007}. \citet{Kuhn.2007} interpret this He I emission as interstellar in origin, while \citet{Molnar.2025} argue for  a terrestrial origin.  The hope of these pioneering experiments was to use the He~I~1.0833~\micron\ diagnostic to study interstellar gas streaming into our solar system --- {\bf if}  that signal could be confidently detected and disambiguated from terrestrial and coronal sources. 

How can emission from near the Sun tell us about interstellar gas?  The big picture is this.  The sun is not orbiting the Galaxy at precisely the same velocity as the very local interstellar medium (ISM) gas.  In the Sun's reference frame, a wind of cold  ($<10^4$K) neutral gas from the Milky Way's interstellar medium flows into our solar system at $\sim 26$~\kms\
($5.3$~AU~yr$^{-1}$).  (\citealt{Mobius.2004}, \citealt{Gombosi.2009}, and the delightful review by \citealt{Lallement.2002}.)
While the heliospheric magnetic field modulates the flux of charged particles from the galaxy, neutral atoms flow through the heliopause, carrying into our solar system information about the local ISM.  Almost all the interstellar H becomes ionized (mostly due to charge exchange with the solar wind, \citealt{Lallement.2002}); about half the interstellar He (from 35\% at solar minimum to 56\% at solar maximum; \citealt{Swaczyna.2022}) becomes ionized, mostly by solar photoionization, with contributions from charge exchange and electron impact.  These interstellar ions  are  termed ``interstellar pickup ions'' when detected by solar wind instruments, as they retain a kinematic signature of their interstellar origin.

Given the high ionization potential of helium, the dimness of our sun in the extreme UV, and the relatively high speed of the ISM wind, about half the interstellar helium remains neutral as it passes through our solar system \citep{Lallement.2002} --- so is not deflected by the Sun's magnetic field.  What happens instead is that the sun gravitationally focuses the neutral ISM, forming a cone of denser gas downwind from the sun, a feature known as the ``neutral Helium focusing cone'' (see \citealt{Mobius.2004} for a review.)
Interstellar helium has been studied in the extreme ultraviolet using backscattered He~I~584\AA\ emission (c.f.\ \citealt{Vallerga.2004}), pickup ions \cite{Gloeckler.2004}, and direct particle detection \citep{Witte.1993, Witte.2004}.

From low Earth orbit, it has not been possible to detect the signal of He~I~1.0833~\micron\ emission from interstellar gas, due to the fact that the Earth's upper atmosphere is a strong emitter of He~I~1.0833~\micron.  \citet{Brammer.2014} report that emission in this line from Earth's upper atmosphere increases the background in the IR channel of Hubble's WFC3 instrument by factors of 3--4 (their Figure 1).
\citet{Kulkarni.2025} summarizes the history of this line's use in astronomy, solar physics, and atmospheric science, reports the detection of atmospheric He I by the SPHEREx mission from low Earth orbit, and warns that emission from Earth's upper atmosphere can stymie efforts to use this line to study the gas of our Milky Way galaxy. \citet{Hui.2026}  present 8 months of He I emission data from SPHEREx, all of it presumably terrestrial.

In this paper, we report the serendipitous detection of diffuse He~I~1.0833~\micron\ line emission with JWST's NIRSpec instrument, from a vantage point far from terrestrial emission sources, and exposure times much longer than eclipse totality.   We detect the high-intensity signature of the neutral helium focusing cone, as well as fainter year-round emission, which we attribute to tiny messengers --- neutral helium atoms from the Galaxy, flowing into our solar system.

\section{Data and Methods}\label{sec:methods}
\subsection{Why NIRSpec}
While JWST's two other near-infrared instruments also presumably detect the He I 1.0833~\micron\  emission line, in this paper we use spectra from the NIRSpec instrument.  NIRSpec is the most heavily used instrument onboard JWST, accounting for about half the observing time. NIRSpec is sensitive to emission at 1.0833~\micron\ when the dispersing element is set to either G140M (medium spectral resolution blue grating, spectral resolution $R \sim 1000$)
or G140H (high resolution blue dispersion grating, $R \sim 3000$).
The blocking filter can be either of the two long pass filters (F070LP or F100LP).
While spectra taken with the NIRSpec prism ($R\sim 100$) also 
show an emission feature at 1.0833~\micron, we do not consider prism spectra in this paper because the medium and high resolution gratings provide much better estimates of line intensity, as well as velocity and linewidth information that cannot be gleaned from prism spectra.

\subsection{Why the fixed slits}
In this paper, we reduce and analyze spectra taken with the NIRSpec fixed slits. Fixed slit spectra are obtained anytime the full NIRSpec detectors are read out. As a result, by harvesting archival fixed slit spectra, we construct a uniform dataset of sky spectra from a large number of JWST observations, regardless of the NIRSpec observing mode that was used (microshutters, integral field, or fixed slit.)

The fixed slits are cut into the metal that separates the quadrants of NIRSpec's microshutter array.  Of the five fixed slits, four are calibrated: S200A1, S200A2, S400A1, and S1600A1, where the names refer to the slit width in milliarcseconds (hereafter abbreviated mas), such that S200A1 is one of two fixed slits with a width of 200 mas. The lengths of the slits are 3.3\arcsec\ for the 200 mas wide slits, 3.8\arcsec\ for the 400 mas wide slit, and 1.6\arcsec\ for the 1600 mas wide slit. What matters for this paper is that the narrowest of the fixed slits (S200A1 and S200A2) will have the highest spectral resolution, more or less identical to that of the microshutters (\citealt{Glidic.2025}; appendix of O'Brien \etal\ in prep.)
Further, the longer lengths of the fixed slits compared to the microshutters (the ratio is about 7:1 for the 200 mas wide slits) means that the fixed slits cover more sky than a single  microshutter, and therefore, measure the background more precisely.\footnote{\url{https://jwst-docs.stsci.edu/jwst-near-infrared-spectrograph/nirspec-instrumentation/nirspec-fixed-slits\#gsc.tab=0}}

In this paper, we analyze data from NIRSpec's  S200A1, S200A2, and S400A1 fixed slits.

\subsection{Data used}
From the JWST MAST archive, we downloaded all publicly available NIRSpec datasets taken with either the G140H or G140M dispersion gratings, and either of the F070LP or F100LP long pass filters, that had effective exposure times greater than 500~s. Experimentation showed that exposure times much shorter than 500~s  were unlikely to have well-measured He I emission.
We drew data from 46 individual observing programs, spanning 3.3 years,  with a total exposure time (header keyword EFFEXPTM) of $\sim 22$~d.

We use metadata such as time and sky position from the JWST headers.  We use JWST's ephemeris from the JPL Horizons database.

\subsection{Data reduction}
We reduced the fixed slit spectra using the JWST pipeline \citep{Bushouse.2025} version 1.20.2, using the reference files recommended for that pipeline version: CRDS context \texttt{jwst\_1464.pmap}. We use the default pipeline parameters, with two exceptions: We use the pipeline's implementation of the NSClean algorithm \citep{Rauscher.2024} to remove, at the stage 2 spectroscopic pipeline step, the correlated read noise from the detector images;  and we set the source type as ``extended'', since we are investigating a diffuse background.

We use the JWST pipeline to create reduced spectra for each individual exposure, known as the ``level 2'' spectra.  Doing so enables measurements at the best possible time resolution.
In general each exposure generates an extracted spectrum for each of the fixed slits.
From the two-dimensional level 2 S2D file for each exposure and each fixed slit, we mask the 5 highest and 5 lowest 5 rows in the spatial dimension, since experimentation shows the reduced spectra often have a ``bathtub'' spatial profile.  We then extract a 1D spectrum by taking the median over the spatial dimension; for the uncertainty we take the median absolute deviation over the spatial dimension.  
From the extracted spectrum, we subtract a linear continuum that is the median intensity within $\pm 0.02$~\micron\ of the expected position of the He I line.

We also use the pipeline to combine multiple exposures to create ``level 3'' or combined spectra.  Such spectra reach fainter flux limits than individual exposures due to the greater exposure time.  We extract spectra from the level 3 S2D files in the same way as for level 2.

For other investigations that use JWST background spectra, it may be important to remove from consideration observations where a discrete source was present in the slit.  We have not done so here, since we are looking for a narrow emission line feature.  Taking the median in the spatial dimension will remove much of the light from any discrete source that may be present, and subtracting the continuum should remove much of the rest.

\subsection{Measuring line intensity, linewidth, and redshift}
The model we fit to the continuum-subtracted 1D spectrum is a single gaussian emission line with zero continuum.  While the feature is a triplet, the resolution of NIRSpec's gratings is insufficient to separate the components; the wavelengths are separated by only $c\lambda/\Delta\lambda\ = -34.8$~\kms\ and $+2.7$~\kms,
which is much smaller than the FWHM of the linespread function (see
Appendix~\ref{sec:appendixLSF} and Table~\ref{tab:R}).
We therefore fit the He I feature with a single gaussian  with a rest-frame vacuum wavelength of 1.083313778~\micron, which is the average wavelength of the multiplet weighted by the statistical weights.
For fitting, we use the python package lmfit, with weights set to $1/\sigma$, where $\sigma$ is the uncertainty spectrum (which we take to be the median absolute deviation in the spatial direction).  We fit the portion of the spectrum within $\pm 0.003$~\micron\ of the expected position of the He I feature.

\begin{figure}
\centering
\includegraphics[width=14cm]{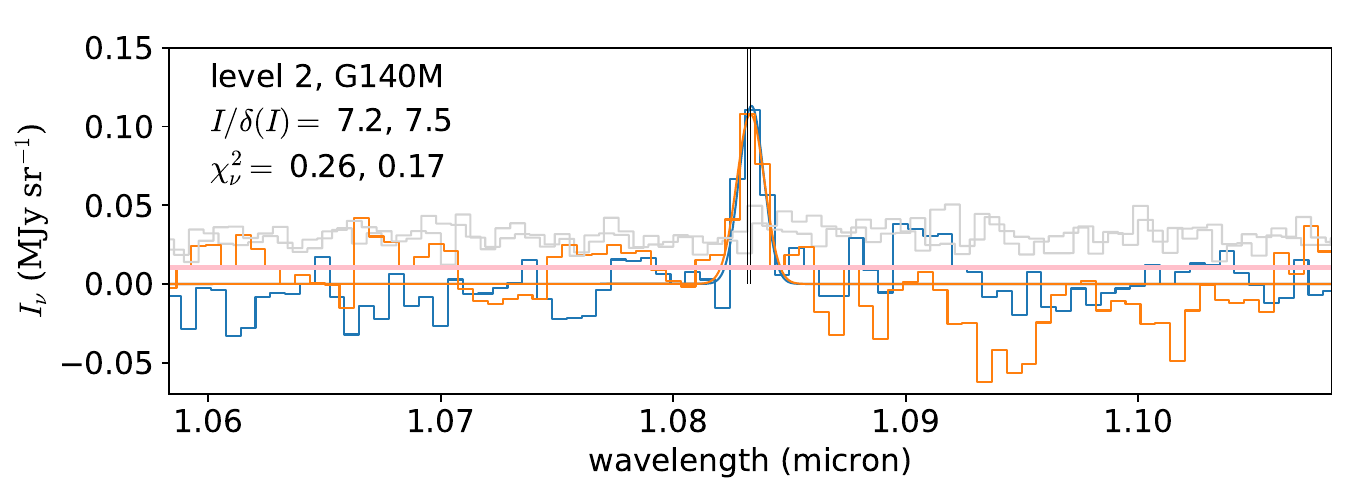}
\includegraphics[width=14cm]{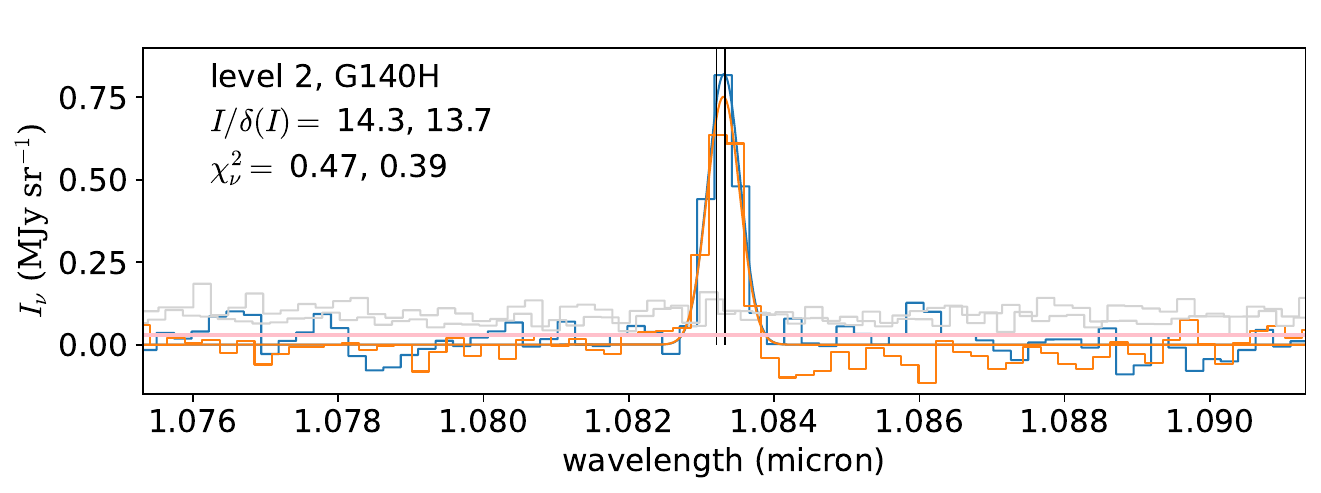}
\caption{Example extracted 1D spectra and gaussian fits to the He~I emission line.  These are exposure-level (level 2) extractions of the two 200 mas wide fixed slits, for the exposures shown in Figure~\ref{fig:skyline}. Each 1D spectrum is plotted as steps, with the best gaussian fit plotted as a smooth curve of matching color. Rest-frame wavelengths of the He I triplet are plotted as vertical black ticks.  The top panel shows G140M (the medium--resolution grating), and the bottom panel shows G140H (the high--resolution grating).  Slit S200A1 is plotted in blue, and S200A2 in orange.} 
\label{fig:1Dspectra}
\end{figure}

We paramaterize the gaussian such that the outputs of the fitting process are the line intensity, linewidth, and redshift, and the uncertainty of each.
The fitter failed for $2.9\%$ of the level 3 spectra, and $3.6\%$ of level 2 spectra; we remove these failed spectra and fits from further consideration. 
See Figure~\ref{fig:1Dspectra} and Appendix Figure~\ref{appendix:examplespectra} for examples of spectra and their gaussian fits.

\subsection{Units}
We report $I$, the radiant intensity of the He I line, also 
known as the surface brightness or the radiant flux per steradian.
We report $I$ in units of \MJysrmicron.
To convert to cgs units (\SBcgsunits ), one should multiply by $0.0025545$.
To convert to units of Rayleigh, one should multiply by $1.751 \times 10^4$ \citep{Kulkarni.2025}.

\subsection{Grouping on the sky}\label{sec:groupingonsky}
We group spectra by sky coordinates, declaring a position to be a new pointing if it is separated from other pointings by more than 15\arcmin .  There are 43 such pointings.

\subsection{Subsamples}\label{sec:subsamples}
We refer to the set of all level 2 extracted spectra with converged fits ($N=3325$) as the full sample.
The total exposure time is $22$~d.\footnote{To avoid over-counting exposure time, we sum exposure time for only the S200A1 slit.}  

To make subsets of higher-quality measurements, we filter by  $I / \delta(I)$, the ratio of measured He I line intensity to its uncertainty as reported by the fitter.
For analyzing line intensity and its variation, we make subsample~A,
requiring  $I / \delta(I) > 8$; this subsample contains  $N=1796$ level 2 spectra
with a total exposure time of $13$~d.\footnote{To avoid over-counting exposure time, we sum exposure time for only the S200A1 slit.}   

For analysis of He I linewidths and redshifts, we make a subsample
with the highest quality and highest spectral resolution, by requiring $I / \delta(I) > 15$, and the narrowest (200 mas) slit width.   We call this subsample~B.  In subsample~B are $N=31$ level 2 spectra taken with the G140H grating (totalling 8.2~hr of exposure time), and $N=194$ spectra taken with the G140M grating (totaling $4.1$~d of exposure time).\footnote{To avoid over-counting exposure time, we compute the summed exposure time for the S200A1 slit, and for the S200A2 slit, and take the average.} 

In Appendix~\ref{appendix:examplespectra}, Figure~\ref{appendix:examplespectra} shows example spectra with $I / \delta(I)$ at the thresholds used to select these subsamples.  From visual inspection, we are confident of detection when $I / \delta(I) \ge 5$.  The conclusions of this paper are insensitive to the choice of threshold levels.

\subsection{Reproducibility}
We make our measurements and relevant metadata available in \S\ref{sec:supportingdata}.
We publish the Jupyter notebooks used to reduce the spectra, analyze the data, and make the plots on Zenodo: \url{10.5281/zenodo.21824563}

\section{Results}
\subsection{Ubiquity of He I emission}
Even though the JWST spectra are heterogeneous in sensitivity (since this is an archival sample with varying integration time and different configurations of grating and blocking filter), we can still examine how frequently the He I line is seen.  We examine the level 3 (combined) spectra, since they are deeper than the spectra from individual exposures.
Of the level 3 spectra,
$89\%$ have  $I / \delta(I) > 5$, 
$80\%$ have  $I / \delta(I) > 8$, and
$55\%$ have  $I / \delta(I) > 15$.
For the level 2 spectra, these percentages are $74\%$, $53\%$, and $22\%$. 
We are reasonably confident that the line is detected when $I / \delta(I) \ge 5$, and quite confident when $\ge 8$.  See  Appendix~\ref{appendix:examplespectra}.  
Thus, despite the heterogeneous nature of the dataset, He I emission is confidently detected in the vast majority of the combined spectra, and in more than half of the individual exposures.

\subsection{Intensity of He I emission}
Table~\ref{tab:flux} reports the He I line intensity we measure from JWST NIRSpec spectra.
The median is 0.00022~\MJysrmicron.  By contrast, SPHEREx detect highly periodic intensities that are higher by factors of 30--200 \citep{Kulkarni.2025}, because SPHEREx is in low Earth orbit, and detects emission from the Earth's thermosphere.

For zodiacal backgrounds at 1.1~\micron\ typical for a dark field like CDF-S ($\sim 0.19$~\MJysr), the He I line intensities we measure with JWST 
correspond to an emission line equivalent width of $\sim$12\AA\ (8--20\AA\ for 25 and 75 percentiles for the full sample). 

\subsection{Variability of He I emission over 3.3~yr}\label{sec:longtermvar}
\begin{figure}
\centering
\includegraphics[width=14cm]{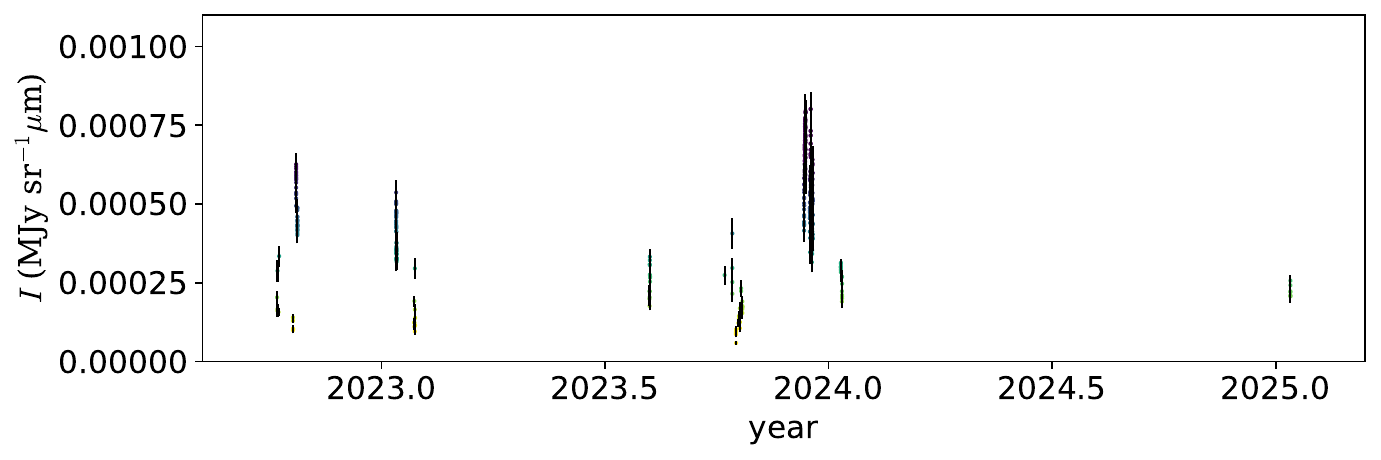}
\caption{Variability with time of He I line intensity over more than two years, for pointings near the Chandra Deep Field South (CDF-S), for spectra with measured $I / \delta(I) > 8$.  The x-axis is decimal year.  Data are from subsample~A.  
} 
\label{fig:varCDFS}
\end{figure}

Famous extragalactic deep fields like the Chandra Deep Field South (CDF-S) have been observed many times by JWST. Figure~\ref{fig:varCDFS} plots the He I 1.0833~\micron\ line intensity versus time for the data in subsample~A taken when the telescope was pointed within 15\arcmin\ of CDF-S.  The He I line intensity is variable from observation to observation, and is also variable {\bf during} some observations, in particular for data taken in late 2023.

\begin{figure}
\centering
\includegraphics[width=\textwidth]{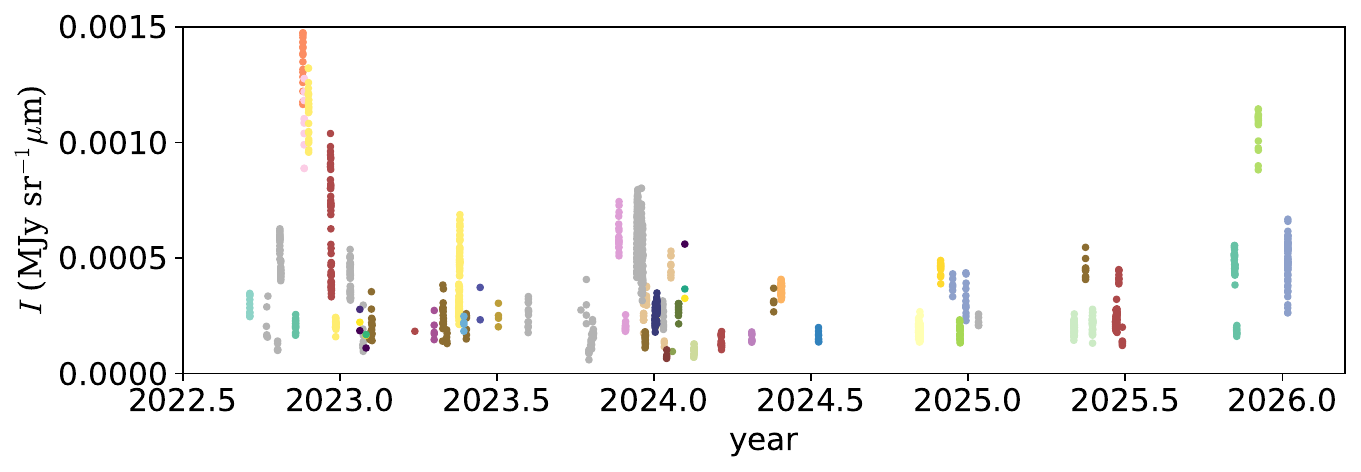}
\caption{Time variability of He I line intensity, with a unique color assigned to each unique sky position.  At a given sky position, the He I line intensity varies by
factors of several on timescales of a day.  Data are from subsample~A.}\label{fig:trending}
\end{figure}

We now explore this variability for all pointings.  
Figure~\ref{fig:trending} plots the He I intensity for all observations in subsample~A as a function of time.  The data are color-coded by position on the sky.  The variation in He I intensity with a given observation is real, not measurement uncertainty.   Clearly, at a given sky position, there is variation in  He I intensity throughout the year, as well as during observations.

\begin{figure}
\centering
\includegraphics[width=\textwidth]{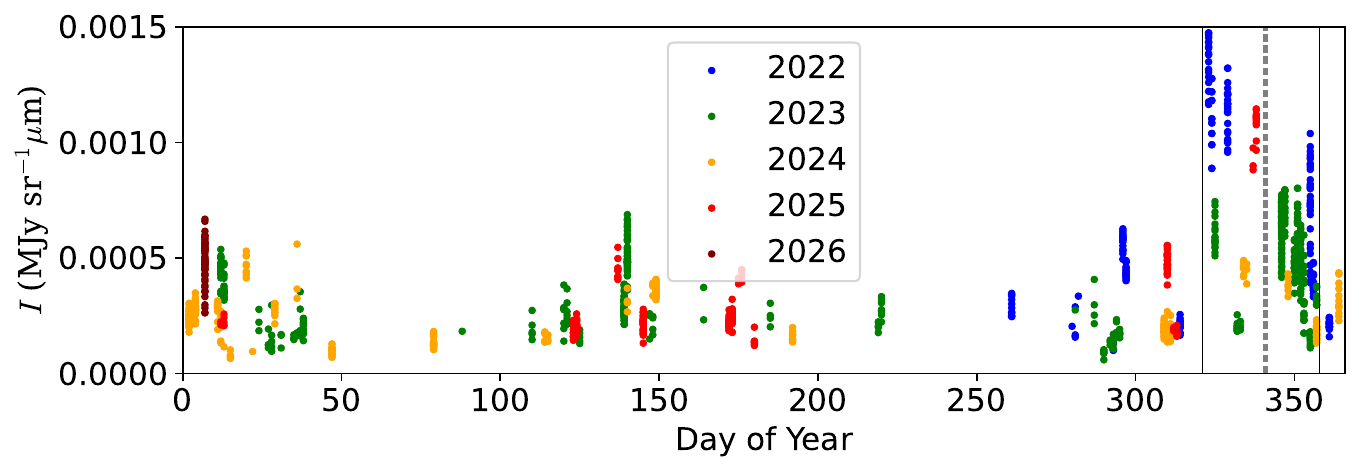}
\caption{He I intensity versus day of year of observation; data are as in Figure~\ref{fig:trending}, but color-coded by year of observation. Black vertical lines mark the times of peak He I intensity, DOY 321 to 358 (Nov.\ 17. to Dec.\ 24.) The grey dashed vertical line marks the date UT Dec.\ 7 00:00:00, when the Earth should cross the center of the helium focusing cone (see Appendix~\ref{sec:appendixconecenter}); the  width of the grey vertical line shows the uncertainty in that predicted time.}
\label{fig:trendingfolded}
\end{figure}

Figure~\ref{fig:trendingfolded} folds the data to highlight the strong seasonal trend.  The x-axis is day-of-year (DOY), and points are color-coded by year.  The He I intensity is highest within a narrow (37 day) window of time, which we manually identify as DOY 321 -- DOY 358, corresponding to Nov.\ 17 -- Dec.\ 24.  This period corresponds to when the Earth (and JWST) pass through the peak density region of the interstellar He focusing cone, as we will discuss in \S\ref{sec:calliscomingfrominsidethehouse}.  Three of the four passages through the He cone have strong He I. 

\begin{figure}
\centering
\includegraphics[width=\textwidth]{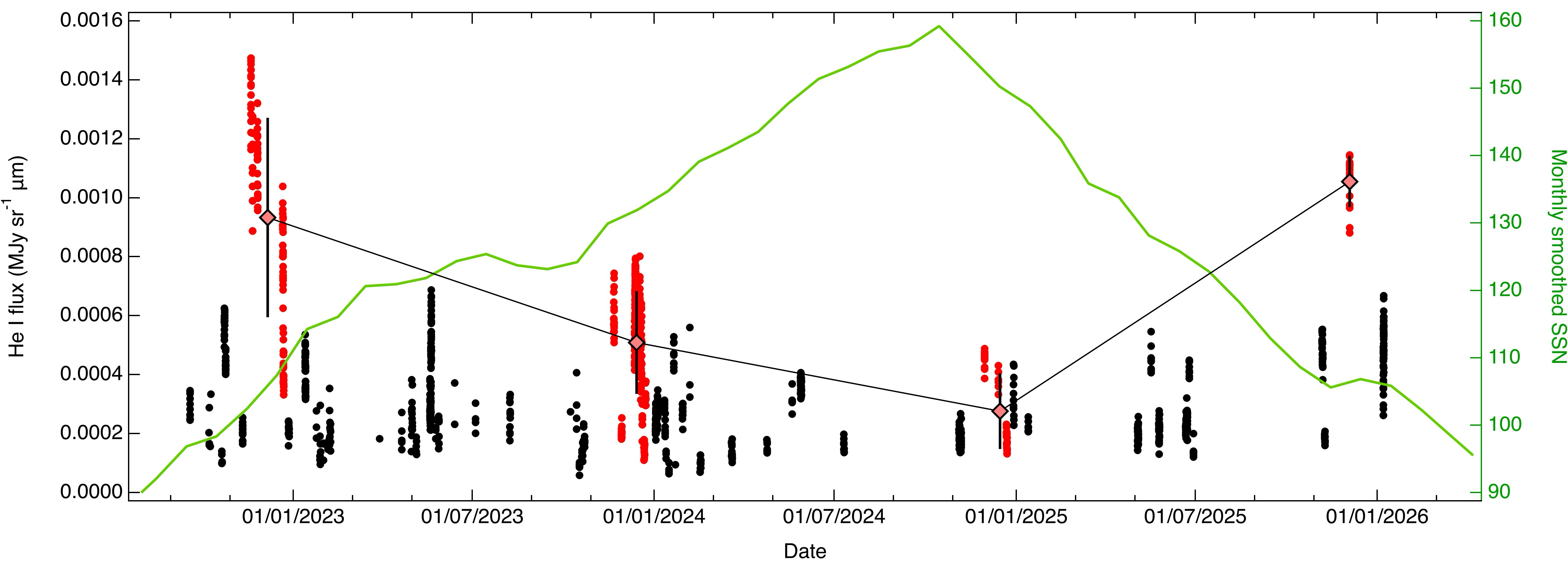}
\caption{He I intensity versus solar activity.  He I data are as in Figure~\ref{fig:trending}, but color-coded red to mark times when JWST was within 20\degr\ of the He I cone center, and otherwise black.  Solar activity, parameterized as the monthly smoothed sunspot number (SSN),\footnote{From \url{https://www.sidc.be/SILSO/INFO/snmtotcsv.php}} is marked by the green curve.  Diamonds show the mean He I intensity for all observations in a given year when JWST was within 20\degr\ of the He I cone center; the errorbars show the standard deviation.  A black line connects the diamonds to guide the eye, and illustrate the anti-correlation with solar activity.}\label{fig:versussolaracitvity}
\end{figure}

Next we compare He I intensity to solar activity.  Figure~\ref{fig:versussolaracitvity} bins observations from subsample~A at times when JWST was within 20\degr\ of the center of the helium focusing cone; the binned points are anti-correlated with the monthly smoothed sunspot number, with a Pearson correlation coefficient of $-0.985$.  This anti-correlation nicely explains why the cone crossing in late 2024 is the only one that lacks strong He I --- it was the crossing closest in time to solar maximum.

\subsection{Short-term variability of He I emission}\label{sec:shorttermvar}
\begin{figure}
\centering
\includegraphics[width=13cm]{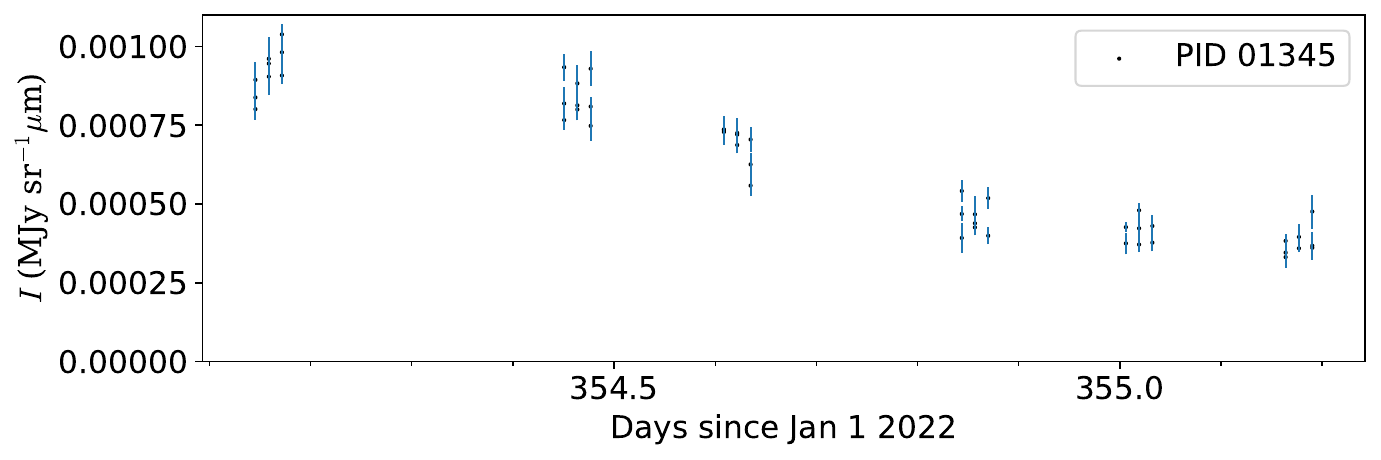}
\includegraphics[width=13cm]{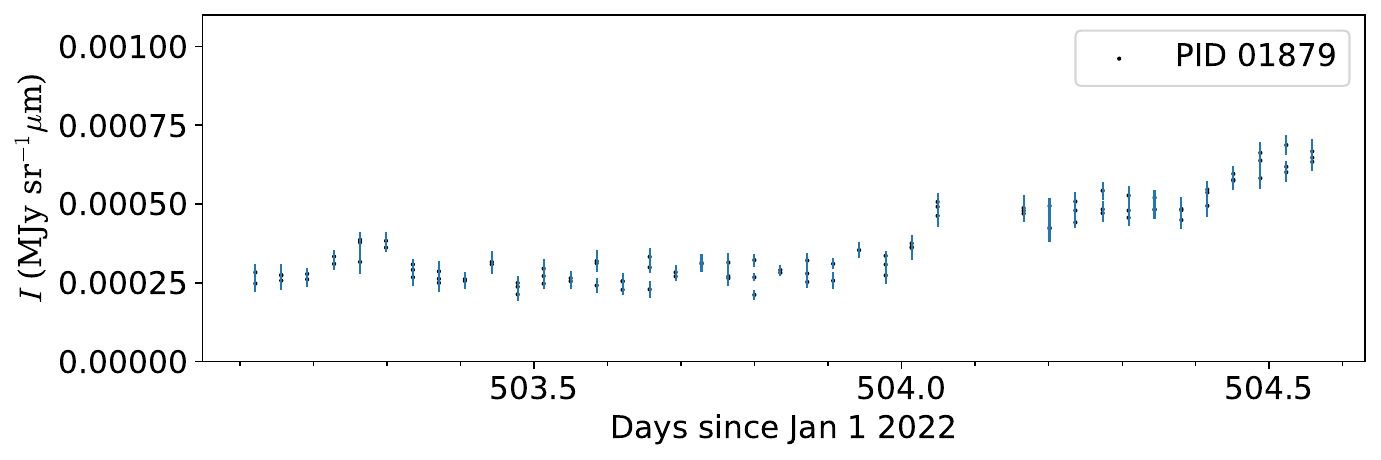}
\caption{Time variability of He I intensity on hour-long timescales, for pointings in the Extended Groth Strip (EGS) (top panel) and COSMOS (bottom panel). Observations in the top panel were obtained 21--23 Dec.\ 2022; observations in the bottom panel were obtained 19--21 May 2023.  In both cases, the He I intensity changes fairly smoothly by a factor of 2--3 over 1--2 days. Data are from subsample~A.
}
\label{fig:morevar}
\end{figure}

The vertical spread in Figure~\ref{fig:varCDFS} and Figure~\ref{fig:trending} reveals considerable variability in He I intensity {\bf during} observations.  Figure~\ref{fig:morevar} zooms in on that variability, showing that during $\sim 1.3$~d long  observations of the Extended Groth Strip (EGS) and COSMOS, the He I intensity changed by a factor of $\sim 2$.
{\bf We are not aware that such extreme short-term variability of interstellar neutral helium has been previously reported.}  \citet{Vallerga.2004} reported variability in  He~I~584\AA\ line emission that did not exceed $50\%$.  Interestingly, we see this high variability both when JWST was inside the neutral helium focusing cone (Figure~\ref{fig:varCDFS} and top panel of Figure~\ref{fig:morevar}), and when it was outside the cone (bottom panel of Figure~\ref{fig:morevar}). 

\subsection{Range of observed He I intensity}
The ratio between the highest intensities seen within the cone, and the typical baseline intensity is about a factor of 6.  This is lower than, but qualitatively similar to, the factor of 13  in He I 584\AA\ emissivity predicted by the model of \citet{Michels.2002}.

\subsection{Linewidth}
\begin{figure}
\centering
\includegraphics[width=\textwidth]{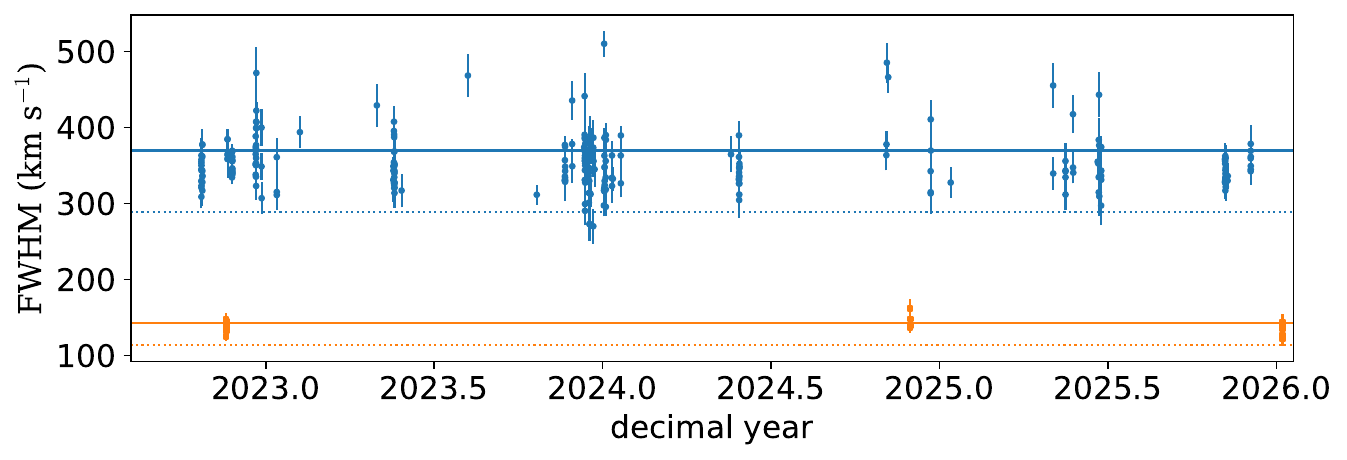}
\caption{Measured linewidth of the He I emission, for subsample~B.  Blue circles
  show measurements made with the G140M grating, and orange squares show
  measurements made with the G140H grating.
  In the same color-coding, for each grating the dotted line shows the FWHM of the point-source LSF (from \citealt{Shajib.2025}, and the solid line shows the LSF for a uniformly-illuminated slit as calculated in Appendix~\ref{sec:appendixLSF}.
  The measured linewidths are very similar to the uniformly--illuminated LSF.}
\label{fig:linewidth}
\end{figure}

Figure~\ref{fig:linewidth} plots  measured He I line widths.  
For the G140H spectra in subsample~B, the measured line width is $139 \pm 6$~\kms\ (FWHM median and median absolute deviation, hereafter MAD).
This closely matches the LSF for a uniformly illuminated, 200 mas wide fixed slit that we calculate in Appendix~\ref{sec:appendixLSF} and tabulate in Table~\ref{tab:R}.
For spectra in subsample~B taken with the G140M grating, the measured line width is
$347 \pm  17$~\kms , again closely matching the LSF for a uniformly illuminated slit  (see Table~\ref{tab:R}).
In fact, the median measured linewidths are 0.6 and 1.2$\sigma$ {\bf less than} their computed LSFs. 
Thus, for the best measured spectra, the He I  linewidths are consistent with being spectrally unresolved, which means the intrinsic velocity dispersion must be much less than the instrumental line spread function.

The equation for the velocity width (FWHM) of thermalized gas in a cloud is
\begin{equation}
v = \sqrt{\frac{8 k \ln(2) T}{m}}.
\end{equation}
For the temperature of He in the solar wind, $T=5.8 \times 10^5$~K (\citealt{Justa.1985}, their Table 3.1), and the mass of a helium atom, the intrinsic velocity width should be $v=81$~\kms . Adding this velocity width in quadrature with the linespread function, for G140H such hot gas should have a measured linewidth of 165~\kms, which is 
15\%\ wider than the LSF.  This exceeds the maximum linewidth measured for a G140H spectrum in  subsample~B.  
We rule out such a high temperature at $4\sigma$.

\subsection{Redshifts}

\begin{figure}
\centering
\includegraphics[width=\textwidth]{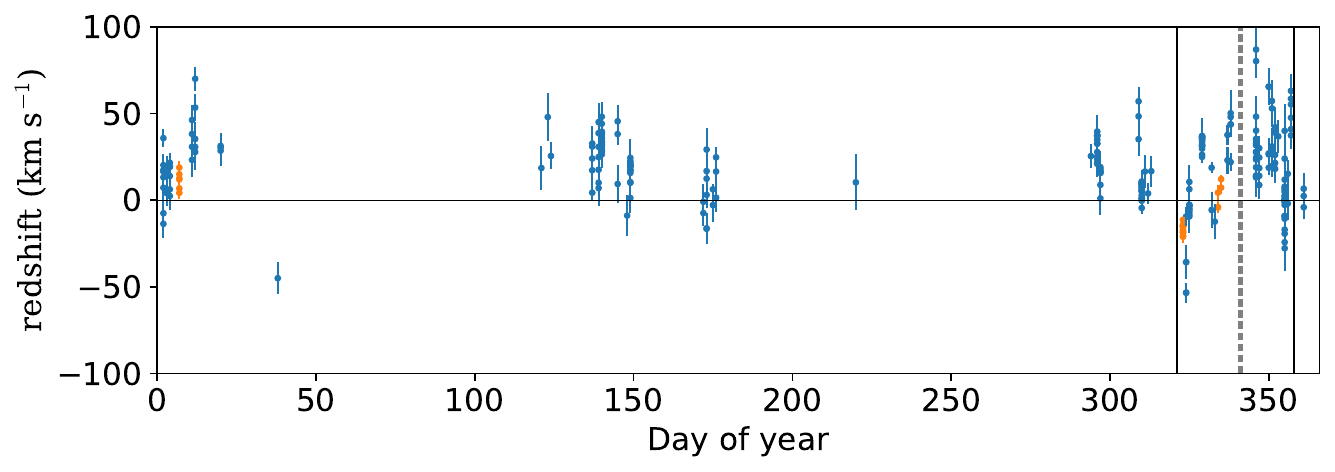}
\caption{Measured redshifts of He I 1.0833~\micron\ emission, as a function of day of year, for subsample~B.  The color-coding is blue for the G140M grating, and orange for the G140H grating.}
\label{fig:redshifts}
\end{figure}

We examine the measured He I redshifts, using these same high-quality, high-spectral-resolution spectra (subsample B).
For G140M, the redshift is       $20 \pm 13$~\kms  (median and MAD).
For G140H, the measurement is   $-11 \pm 23$~\kms , which is limited by the small number  ($N=23$) of high-quality spectra.
Combining the G140H and G140M subsamples, the measured redshift is $19 \pm 14$~\kms .  
See Figure~\ref{fig:redshifts}.

We find no correlation between the measured line redshift and the barycentric velocity correction (the component of JWST's velocity relative to the barycenter of the solar system, in the direction JWST was pointed). Since the JWST pipeline applies the barycentric correction to all spectra, these quantities should appear correlated in a scenario in which JWST itself is the source of the Helium emission. Such a scenario would also predict a seasonal variation in Figure~\ref{fig:redshifts}. Neither effect is observed.

\begin{figure}
\centering
\includegraphics[width=\textwidth]{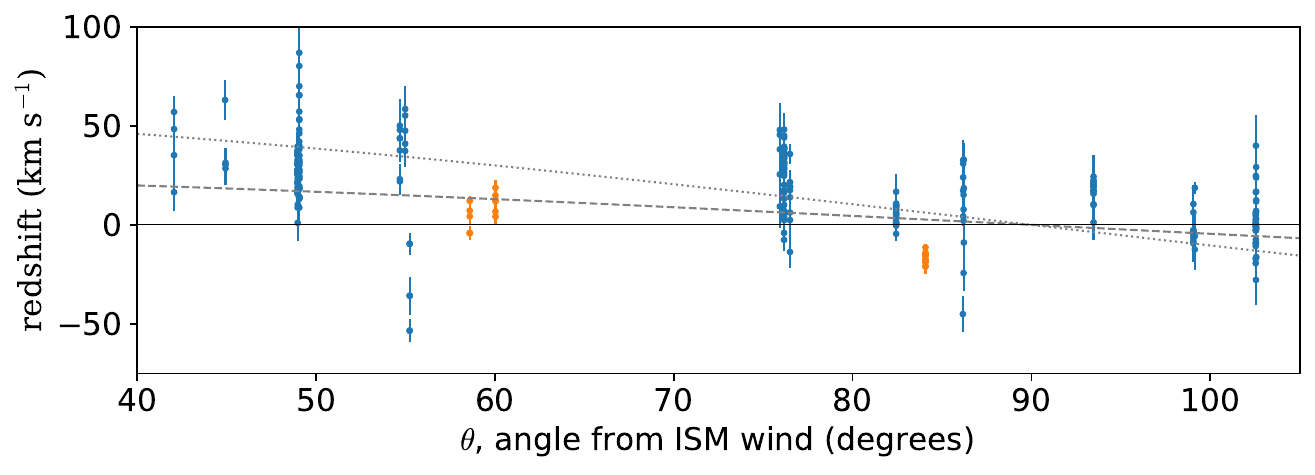}
\caption{Measured He I redshift as a function of the ISM wind angle (defined as the angle between the telescope's pointing direction and the wind direction as given in Appendix~\ref{sec:appendixconecenter}.)  The data subsample and color-coding are the same as in the previous figure.
  The grey dashed line indicates  $26\cos(\theta)$~\kms\ (the expected behavior of the interstellar wind far from the sun) and  $60\cos(\theta)$~\kms\ (typical of gas that has been deflected by the sun.}\label{fig:redshiftvwindangle}
\end{figure}

The sun is moving with respect to the very local interstellar medium at a speed of 26~\kms . This creates an apparent ``ISM wind'' in the direction given in Appendix~\ref{sec:appendixconecenter}. 
We now test whether this wind can explain the measured redshifts. For each observation, we calculate the angle between this wind direction and where JWST was pointed. For subsample~B, the range of angles is 42\degr\ to 103\degr, with a median of 76\degr . See Figure~\ref{fig:redshiftvwindangle}. These are not ideal angles for measuring radial velocity --- no pointings are close to pure upwind or downwind.
We consider only those spectra in subsample~B with wind angle $\theta$  $< 70$\degr\ or $>110$\degr : for these spectra,  the measured  wind speed,
corrected by $1/cos(\theta)$, is $42 \pm 16$~\kms\ (median and MAD).
Within uncertainties, this is consistent with the expected $26$~\kms\ speed of the ISM wind.

As it is deflected by the sun, interstellar neutral helium is accelerated to 50--80~\kms\ at Earth orbit \citep{Starkey.2025}.  The measured redshifts should therefore be higher when JWST is within the focusing cone.
Using the same wind angles as above, we measure an absolute value of wind speed corrected by $1/cos(\theta)$ of
$46 \pm 17$~\kms\ within the cone, 
$43 \pm 30$~\kms\ for the rest of the year.   
The sample of high-quality spectra is simply too small to determine whether the redshift is different inside of the cone. 

\subsection{No dependence on solar angle}

\begin{figure}
\centering
\includegraphics[width=\textwidth]{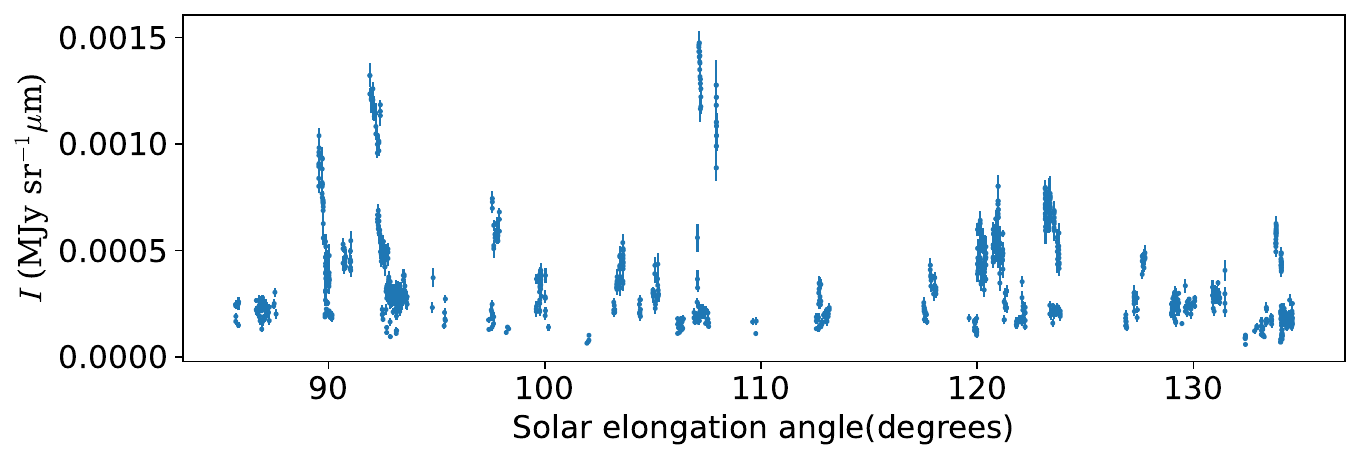}
\includegraphics[width=\textwidth]{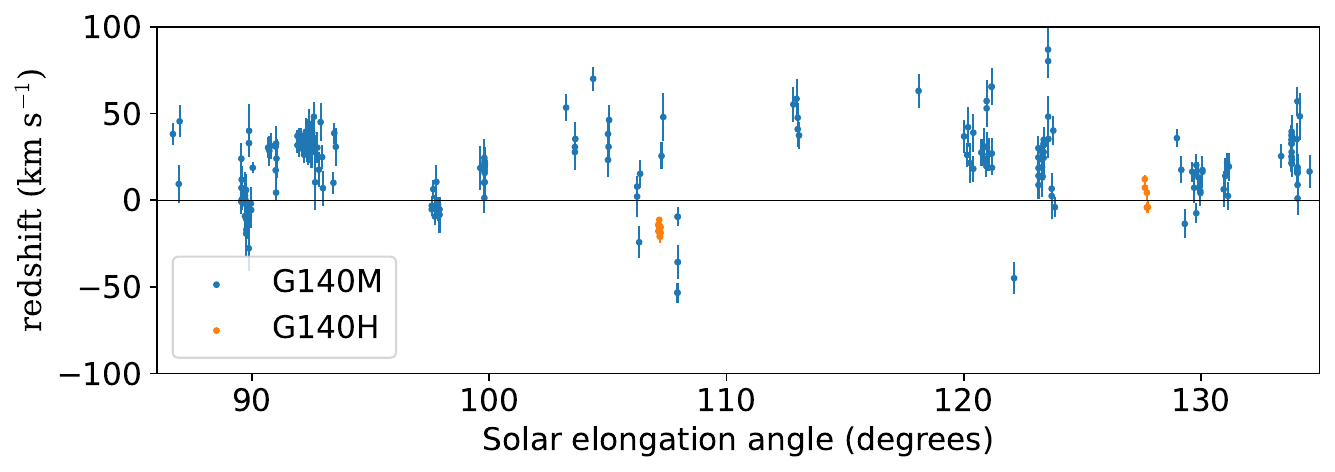}
\caption{No correlation between solar elongation angle and He I 1.0833~\micron\ intensity or redshift. Top panel: Y axis is He I line intensity; X axis is the angle between where JWST was pointed and the sun.  Data are subsample~A.
  Bottom panel: same X axis; Y axis is the redshift of the helium line.
Data are subsample~B, color-coded as in the previous figure.}
\label{fig:solarelongangle}
\end{figure}

We see no correlation of either He I intensity or He I redshift with the solar elongation angle (the angle between where JWST was pointed and the sun.) See Figure~\ref{fig:solarelongangle}.
Such correlations would be expected if the sun were the source of the observed neutral helium.

\section{Discussion}

\subsection{The neutral helium cannot arise in the solar wind}\label{sec:notsolarwind}

The properties of the He I emission are incompatible with an origin in the solar wind, for the following reasons. While He is indeed abundant in the solar wind, comprising $\sim$5\% of the particles (Hargreaves, chapter 5.3.2), the He is overwhelmingly doubly ionized \citep{Justa.1985}, not neutral. This ionized He has typical speeds of $\sim$400~\kms\ (\citealt{Justa.1985}, their table 3.1); if what we were seeing were alpha particles that have doubly--recombined, they should still be moving fast.  In addition, the hot particles of the solar wind (alpha particle $T=6 \times 10^5$~K, \citealt{Justa.1985} table 3.1) should produce linewidths that are broader than the instrumental line function.  Contradicting this model, we measure a low redshift and rule out such broad linewidths at $4\sigma$.
\begin{figure}
\centering
\includegraphics[width=14cm]{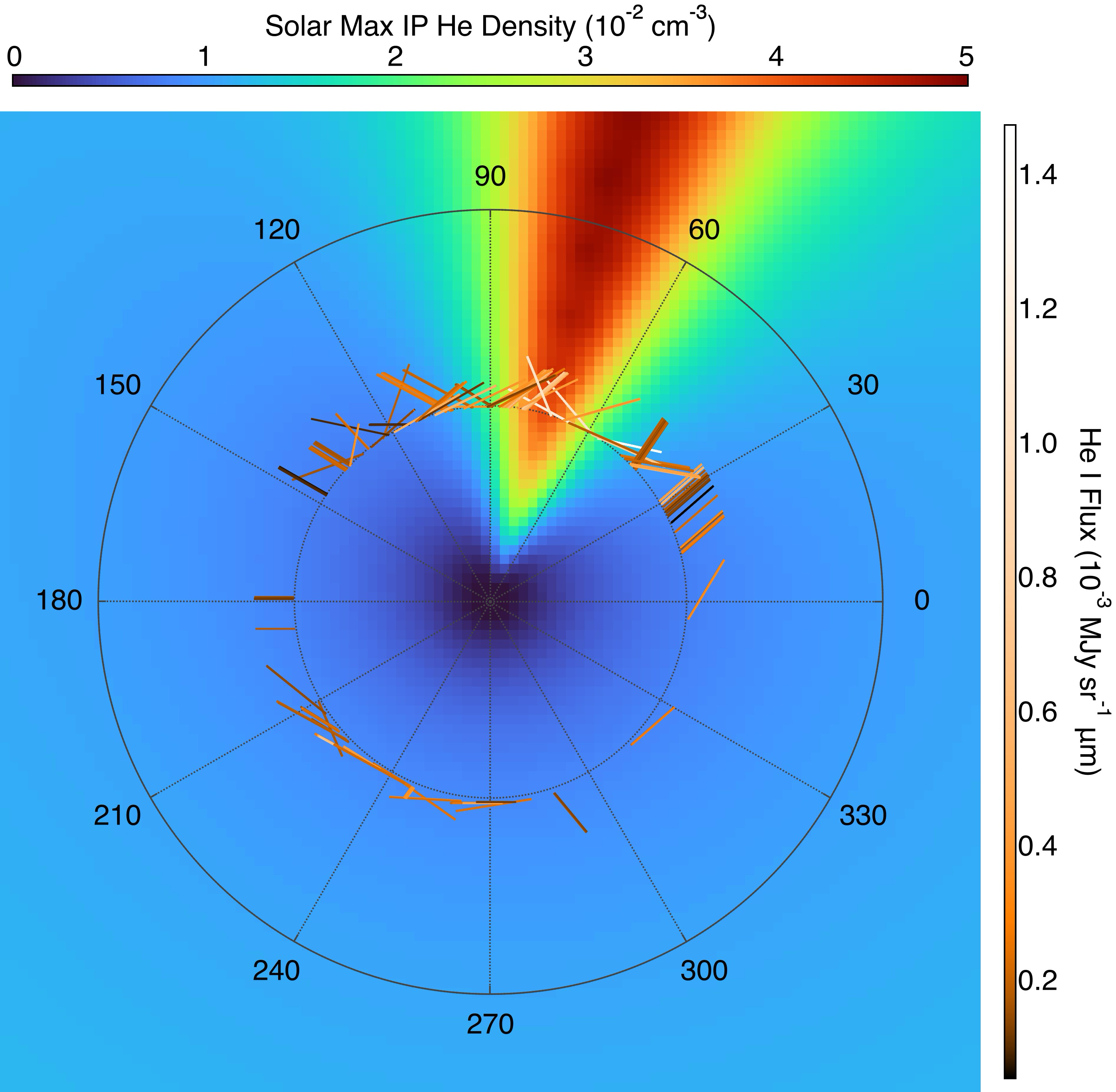}
\caption{The neutral He density within the ecliptic plane, in units of cm$^{-3}$ (top colorbar).  The Sun is in the center; the dotted and full circles correspond to radial distances of 1~AU and 2~AU; ecliptic longitude is marked.  Short lines show JWST's position during each observation and its line of sight projected into the ecliptic plane; they are color-coded for He I intensity (colorbar at right).   The neutral helium focusing cone is the pronounced density enhancement at the top. Model from \citet{Koutroumpa.2009}, based on \citet{Lallement.2004} and \citet{Daulaudier.1984}. JWST's excursion out of the ecliptic plane is not shown, but it is small:   a maximum of 0.003~AU, compared to the $\sim0.5$~AU extent of the He cone.}\label{fig:studythediagramdougal}
\end{figure}

\begin{figure}
\centering
\includegraphics[width=10cm]{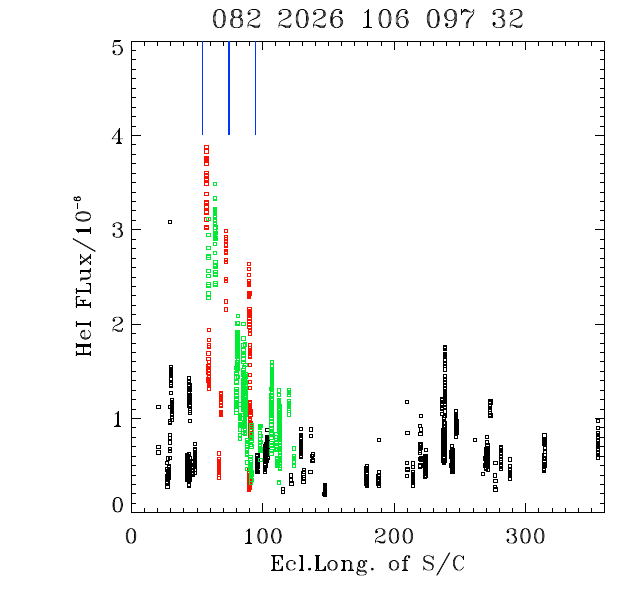}
\caption{He I intensity (in cgs units) as a function of spacecraft ecliptic longitude (in \degr).   Blue lines and red symbols indicate ecliptic longitudes for which the spacecraft is taken to be in the helium focusing cone (within 20\degr\ of the center of the cone as measured from the sun).
Green symbols mark observations where the line of sight was pointed toward the helium focusing cone. }\label{fig:fluxinsccords}
\end{figure}

\subsection{Does the neutral helium emission come from interstellar pickup ions?}
We explore the hypothesis that the He I emission is primarily due to  interstellar pickup ions:
He atoms that enter the solar system neutral and become ionized by EUV solar photons and electron impact.  These newly created interstellar pickup ions could subsequently charge-exchange with neutral H and He in interplanetary space.
This scenario would explain the annual variability of the He I emission intensity, since the pickup ion distribution is also roughly aligned with the interstellar He cone \citep{Gloeckler.2004}.  Moreover, the pickup ion population due to electron impact is much more sensitive to short-term variability of the solar wind, in particular density enhancements, which could explain the observed short-term variability of He I intensity.

The dependence of the pickup ion densities with the solar cycle, however, is a less trivial matter to grasp, because of local effects on the pickup ion production. According to previous data analyses and theoretical models, pickup ion fluxes at $\la 1$--$1.5$~AU are anti-correlated with the solar cycle (just like neutral He density), while at larger distances they are globally correlated with increasing solar activity (Figure~1b of \citealt{Gloeckler.2004}; Figure~10 of \citealt{Lallement.2004}; Figures 14--16 of \citealt{Rucinski.2003}). Since the He I line intensity observed by a spacecraft is the integration of the local emissivity along the line of sight, its anti-correlation to the solar cycle may be highly non-linear, if the pickup ions were the origin of the emission, and difficult to reconcile with the anti-correlation shown in Figure~\ref{fig:versussolaracitvity}.

Another aspect that undermines this scenario is the redshift of the He I line. As they are produced, pickup ions are swiftly picked-up and start gyrating around the interplanetary magnetic field lines and expand radially with the solar wind. This results in a rapid acceleration of the particles to near or higher than solar wind velocities ($>200$~\kms; \citealt{Zirnstein.2022}), which does not agree with the small redshifts observed by JWST.

\subsection{The local ISM as the source of the neutral helium}
\label{sec:calliscomingfrominsidethehouse}

We  conclude that the neutral helium is interstellar in origin, for three reasons.  First, dramatic peaks in He I line intensity all occur within the narrow seasonal window from Nov.\ 17 to Dec.\ 24.  Three of the four years with data exhibit flares of emission in that narrow window. This window, centered on early December, corresponds to the time of year when the Earth passes through the neutral Helium focusing cone.
Figure~\ref{fig:studythediagramdougal} shows a schematic of this helium focusing cone within the ecliptic plane.
In Appendix~\ref{sec:appendixconecenter}, we calculate when the Earth passes through the center of the helium focusing cone; this time, plotted as the grey dashed line in Figure~\ref{fig:trendingfolded}, lies within 2 days of the center of the window of high He I intensity that we had marked by hand. That the periods of highest detected He I intensity occur annually, in a 37--day-wide window centered within 2 days on when L2 passes through the He focusing cone, is strong evidence that JWST's He I 1.0833~\micron\ background is interstellar in origin.

In Figure~\ref{fig:fluxinsccords} we expand on this point, marking observations taken when the  spacecraft was within 20\degr\ (as measured from the sun) of the center of the helium focusing cone, as well as observations when the line of sight was pointed toward the helium focusing cone.  In both cases, the He I intensity is elevated.

The observed anti-correlation between He I intensity within the focusing cone and solar activity is also neatly explained by this model, since interstellar helium is more likely to be ionized (and thus unable to produce He I emission) by the increased photoionization and electron impacts associated with solar maximum.  

Second, we measure a  low median redshift for the He I emission
($19 \pm 14$~\kms) .  
Correcting for $1/cos(\theta)$, and ignoring pointings orthogonal to the wind, this implies a wind speed of $42 \pm 16$~\kms .  
These measured redshifts are consistent within $1\sigma$ with the expected 26~\kms\ speed of the ISM wind,
especially when we consider that the wind speed should be higher within the He focusing cone due to gravitational acceleration by the sun \citep{Starkey.2025}.

Third, the linewidth of the feature is narrow, consistent with the line spread function, which is what is expected for the cold gas of the ISM. 

Therefore, we identify the ubiquitous diffuse 1.0833~\micron\ line emission seen in JWST NIRSpec spectra as arising from neutral helium in the interstellar medium, and we identify the periods of elevated line intensity as caused by JWST passing through the cone of this helium that is gravitationally focused by the Sun.

\subsection{Impact on measuring the MW halo}
\label{sec:notMWHalo}
\citet{Kulkarni.2024} proposed that the He I 1.0833~\micron\ line be used to determine the ionization state of the Milky Way halo, adapting the method of  \citet{Reynolds.1995}. Indeed, that intriguing possibility motivated our initial investigation.  Regrettably, given the evidence for a bright, variable interstellar foreground presented in this paper, it seems  unlikely that He I~1.0833~\micron\ can be used to measure the ionization state of the outer Milky Way.

\subsection{Impact on JWST observations}
We now consider whether the time-variable He I 1.0833~\micron\ background has any impact on JWST observations or observation planning.

The equivalent widths we measure (8--20\AA\ for 25 and 75 percentiles for the full sample; see Table~\ref{tab:flux}) mean that the total background will be several times more than the zodiacal background at the position of the He I line. Practically speaking, JWST has a sky line.  Even when the sky background is carefully subtracted from the target spectrum, as is typically done, the variability of the He I line, and its high intensity, mean that residuals from background subtraction will be high at 1.0833~\micron .
Given the narrowness of this feature and the breadth of JWST's wavelength range (the He I feature affects $<1\%$ of the wavelength range of NIRSpec's dispersers), for most observations the impact will be negligible. 

The He I 1.0833~\micron\ spectral feature is also used to study exoplanet atmospheres, primarily with the NIRISS instrument in slitless spectroscopy mode  \citep{Louie.2025, Gressier.2025, Allart.2025}.  We briefly consider the possible impact on these observations of the variable, diffuse He I foreground.   We  do not believe that {\bf slitless} observations (i.e., with the single object slitless spectroscopy (SOSS) mode of the NIRISS instrument, or the short wavelength grism time series mode of the NIRCam instrument) will be affected, for the following reason. The line spread function of the He I foreground will be extremely broad, since this diffuse emission comes in from the entire entrance aperture.  Thus, the He I foreground will be blurred into a smooth background.   By contrast,  He~I~1.0833~\micron\ from the targeted planet will have the much narrower line spread function of the point source that is its host star.

Observations using a slit, such as with NIRSpec, may be more subject to contamination by diffuse He I. Depending on the selected slit width, the He I foreground may have a more or less similar line spread function to He I from the exoplanet. The typical choice of the S1600A1 slit for NIRSpec time series observations would result in diffuse He I having a line spread function several times wider than the point source LSF for star+exoplanet He I (compare the black and blue curves in Appendix~\ref{sec:appendixLSF}, Figure~\ref{fig:LSF}); in principle the different LSFs should help disambiguate any potential contamination. Users planning such slitted observations, particularly if using a narrower slit, should model using the appropriate extended-source LSF (Appendix~\ref{sec:appendixLSF}) the effect of a time-variable diffuse He I foreground, which may be significant for fainter target stars.

\section{Directions for further investigation}\label{sec:futurework}

\subsection{Modeling the He I intensity}
Reconciling the JWST He I intensity measurements to models of the structure of the focusing cone and the dynamic solar environment is beyond the scope of this paper. Future efforts toward that goal will require determining how the 1.0833~\micron\ photons are generated, and how helium is excited to the metastable state so that it is able to resonantly scatter the 1.0833~\micron\ photons.  We briefly explore these issues and present a toy model to stimulate discussion.

Studies of He~I~1.0833~\micron\ in the Earth's upper atmosphere assumed that the Sun was the source of the 1.0833~micron\ photons \citep{Bishop.1993, Kulkarni.2025}.  In solar spectra, He~I~1.0833~micron\ appears in absorption; the absorption is deeper in active regions where UV photons over-populate the metastable state; the absorption is weaker in coronal holes where the UV flux is lower \citep{Son.2021}.  However, He~I~1.0833~\micron\ has been observed in emission during solar flares \citep{Kuckein.2015}.  Its formation in the chromosphere depends on the local conditions (EUV excitation/ionization, electron temperature and density), so it is a tracer for many regions in the chromosphere and corona \citep{Leenaarts.2016}.  Understanding the diffuse He~I emission within our solar system will require estimates of the sun's production of 1.0833~\micron\ photons with time resolution of hours.

Also needed is an understanding of the excitation mechanism --- how is the metastable state  populated?  The interaction energy must be in the range $19.8 < E < 24.58$~eV, to populate the level without ionizing the helium atom.  Possibilities include
photo-excitation from far-UV solar radiation,  as is the case for He I 1.0833~\micron\ emission in the Earth's atmosphere \citep{Kaifler.2002,  Kulkarni.2025};
and collisions with electrons from the solar wind.

\begin{figure}
\centering
\includegraphics[width=\textwidth]{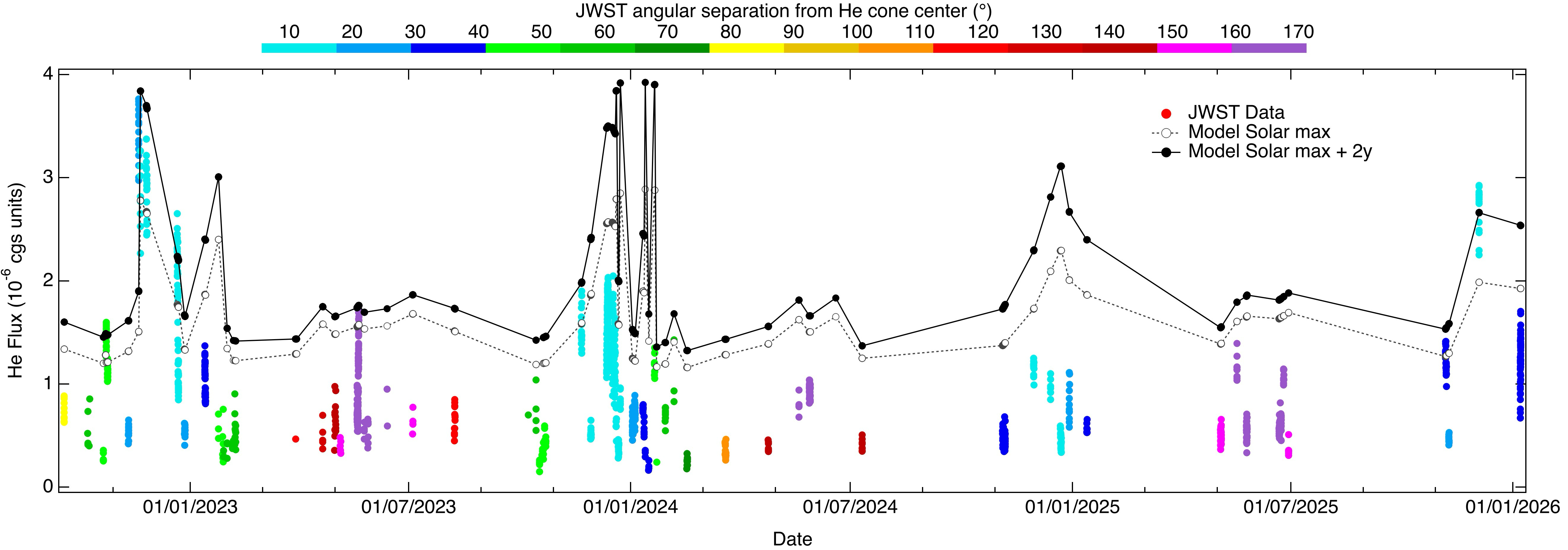}
\caption{Comparison of the data with a very simple model (black points, described in \S\ref{sec:futurework}) that predicts He~I emission for each of JWST's sightlines, assuming that 1.0833~\micron\ photons from the sun are resonantly scattered by neutral He from the ISM wind.  The JWST data are as in Figure~\ref{fig:trending}, but now color-coded according to the spacecraft's angular separation from the He cone center.  The model predictions are connected with straight lines to guide the eye.}\label{fig:toymodel}
\end{figure}

While detailed modeling of these and other effects is beyond the scope of this paper, in Figure~\ref{fig:toymodel} we present a very simple  model for resonant scattering of solar 1.0833~\micron\ photons, based on neutral He column densities calculated with a kinetic hot model (\citealt{Koutroumpa.2009}, based on \citealt{Lallement.2004} and \citealt{Daulaudier.1984}). Neutral column densities are simulated for two different periods of solar activity: solar maximum peak (roughly year 2024) and solar maximum $\pm 2$~yr, to approximate the time span of the JWST data.  We convert the column densities into He I intensity, assuming a resonant scattering cross-section of $2.5 \times 10^{-16}$ cm$^2$ \citep{Kaifler.2002} and a solar He I flux that decreases as the inverse square of the distance from the Sun, $F_s$(1~AU)$/r^2$, where $F_s$(1~AU) $=$ 0.348~\cgsflux\ is the solar flux at 1 AU (from \citealt{Kulkarni.2025} Figure 6, assuming a solar irradiance of 0.58 W~m$^{-2}$~nm$^{-1}$ and a line width of 0.0006~nm).
This simple model predicts He~I~1.0833~\micron\ intensities of the same order of magnitude as what is measured, and captures some of the behavior of the cone--crossing peaks. For this back-of-the-envelope calculation, we assumed that all of the interplanetary He is available to scatter 1.0833~\micron\ photons; we did not attempt to calculate the fraction of  He in the metastable state, which depends on local excitation conditions such as the solar EUV flux and electron collision rate.  Clearly, more detailed modeling is warranted.

\subsection{Variability}
The biggest surprise from this investigation is the variability of the He I emission on timescales of hours (Figure~\ref{fig:morevar}).  Explaining this variability is a key challenge for future work, especially since neutral helium is presumed to be smoothly distributed in interplanetary space, and slowly evolving through the solar cycle. Short term variations may be linked to solar illumination changes, but these are not expected to be more than $50\%$ of the average conditions, as is also the case in the UV resonant line \citep{Vallerga.2004}.  Solar wind electrons may be important, as in some observations
(notably the one shown in the lower panel of Figure~\ref{fig:morevar}), the variability appears to be correlated with the solar wind temperature.  

\subsection{Uniqueness of the JWST dataset}
Given its position in deep space, JWST can study He~I free of contamination from Earth's atmosphere.  Given its excellent sensitivity, JWST is able to study this He~I emission not only within the focusing cone when densities are highest, but throughout the year.  Most uniquely, JWST's measurements of interstellar neutral He are on much smaller scales than has been previously measured:  the NIRSpec fixed slits cover an angular area of a few square arcseconds, as contrasted with direct particle detection and  He~I~584\AA\ backscatter measurements, which have beam sizes of degrees.    As such, the measurements reported here are extremely narrow pencil beams that probe gas on small spatial scales not previously studied.  It is possible that this difference in beam size is related to the extreme variability we observe.  JWST captures this emission on small angular scales and short timescales, opening up a new way to explore the solar environment and the very local ISM. 

\subsection{Closing thought}
As JWST continues to observe the distant universe with NIRSpec using the G140M or G140H dispersers, additional measurements of the neutral helium wind will accumulate.  This paper demonstrates the usefulness of passively gathering background measurements from the NIRSpec fixed slits.  There are doubtless other scientific problems that can be addressed through innovative use of the spectral archive, an archive that will grow as JWST continues to study the universe.

\begin{acknowledgments}

JR acknowledges support from NASA for personal astrophysical research, which is part of their duties as Senior Project Scientist for JWST.
JR thanks the library at the Space Telescope Science Institute, where browsing through the stacks inspired \S\ref{sec:notsolarwind}; the now-virtual library at the NASA Goddard Space Flight Center, whose librarians tracked down obscure references; and the public libraries of Carroll County and Cecil County in Maryland, USA, which provided quiet places to think.  
JR acknowledges conversations with colleagues Harry Ferguson and Rob Petre that shaped this work at key junctures.  RO and YR were supported by NASA under award number 80GSFC24M0006.
DK acknowledges financial support from the Centre national d’études spatiales (CNES), France
(ROR: \url{https://ror.org/04h1h0y33}) within the framework of the JWST mission.

This work is based on observations made with the NASA/ESA/CSA James Webb Space Telescope. The data were obtained from the Mikulski Archive for Space Telescopes at the Space Telescope Science Institute, which is operated by the Association of Universities for Research in Astronomy, Inc., under NASA contract NAS 5-03127 for JWST. These observations are associated with the following program numbers:  01180, 01181, 01207, 01210, 01212, 01219, 01222, 01231, 01286, 01287, 01292, 01324, 01345, 01810, 01869, 01879, 01914, 02110, 02136, 02344, 02478, 02767, 02784, 02969, 03117, 03215, 03543, 03786, 03788, 03843, 04287, 04446, 04713, 04750, 04762, 05629, 05943, 06591, 06642, 06809, 06811, 09214, 09448, 12468, 12494, and 12497.
\end{acknowledgments}

\section*{Supporting Data}\label{sec:supportingdata}
To support future modeling of the He I~1.0833~\micron\ emission, we publish extensive electronic data products:
\begin{itemize}
\item We publish a machine readable table of our measurements of He I 1.0833~\micron\ emission, with accompanying metadata.   {\bf Data Editor: We will supply a CSV file}
\item We publish our reduced, extracted one-dimensional level 2 and level 3 spectra. {\bf Data Editor: we request advice/assistance as to format} -- the spectra are in a pickle of pandas dataframes at present.
\end{itemize}

\section{Software}

This work made use of the following scientific software:
\software{
the jwst pipeline \citep{Bushouse.2025}, 
Astropy \citep{astropy:2013,astropy:2022}, 
Python \citep{python:software}, and
JSkyCalc \citep{skycalc},  
IDL, particularly the NASA--hosted IDL Astronomy Users Library,
JPL Horizons \citep{JPLHorizons}, 
Jupyter \citep{jupyter:2016}, 
lmfit \citep{lmfit:2016},
Matplotlib \citep{matplotlib:2007}, 
NumPy \citep{numpy:2020}, 
pandas \citep{pandas:2010,pandas:software},
pysiaf \citep{johannes_sahlmann_2019_3516964},  and
SAOImage DS9 \citep{ds9:ascl}.
We used generative AI (Codex, \citealt{codex:2026}) to convert the manuscript, tables, and references into the \LaTeX\ and B\kern-.05em{\sc i\kern-.025em b}\kern-.08em\TeX\ formats.}

\begin{deluxetable}{lccc}
\tablecaption{Measured He I 1.0833~\micron\ intensity\label{tab:flux}}
\tablehead{
\colhead{percentile} &  \colhead{all spectra} & \colhead{$I/\delta(I) >8$}  & \colhead{$I/\delta(I) >15$}}
\startdata
\cutinhead{Level 2 (exposure-level) spectra}  
25th           &  0.00015    &  0.00021    &  0.00028\\
50th  (median) &  0.00022    &  0.00032    &  0.00046\\
75th           &  0.00039    &  0.00051    &  0.00063\\
\cutinhead{Level 3 (combined) spectra}  
25th           &  0.00015    &   0.00017   &    0.00033\\
50th  (median) &  0.00022    &   0.00024   &    0.00049\\
75th           &  0.00040    &   0.00043   &    0.0014\\
\enddata
\tablecomments{Measured He I 1.0833~\micron\ intensity, in units of \MJysrmicron .
  Statistics are given for the level 2 (exposure-level) spectra, and for the (combined) level 3 spectra.
}
\end{deluxetable}

\bibliography{newbib3}{}
\bibliographystyle{aasjournal}

\appendix

\section{Typical quality of spectra}\label{appendix:examplespectra}
In this appendix, to show the typical quality of the data, we plot spectra where He I emission was detected at the threshold levels used in this paper.  Figure~shows examples of the level 2 (exposure-level) data, and Figure~\ref{fig:1DspectraSNRlevel3} shows examples of the level 3 spectra (where all exposures from an observation have been combined.)

\begin{figure}
\centering
\includegraphics[width=8.5cm]{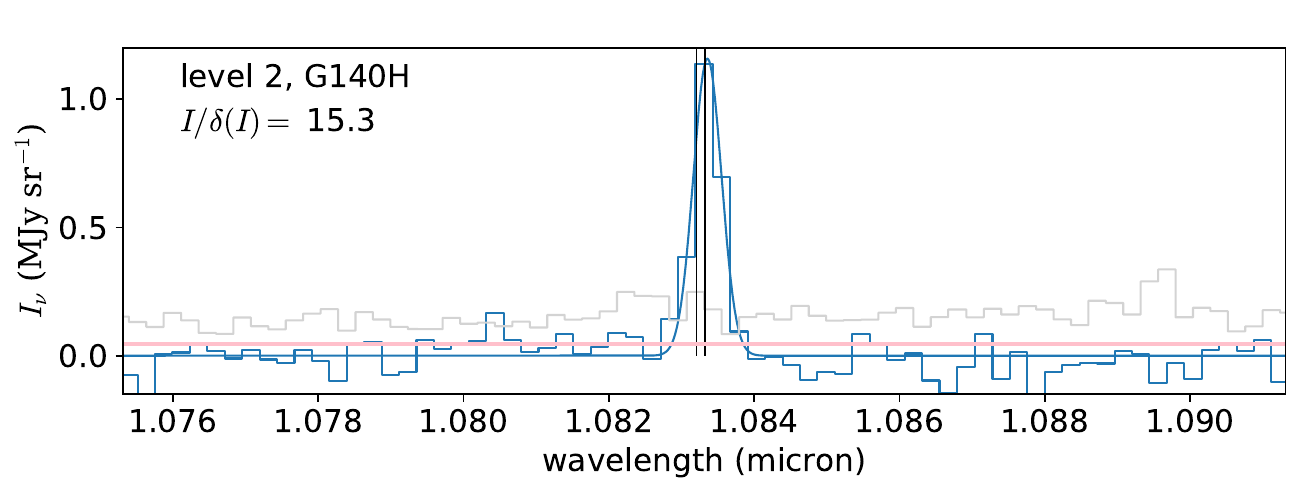}
\includegraphics[width=8.5cm]{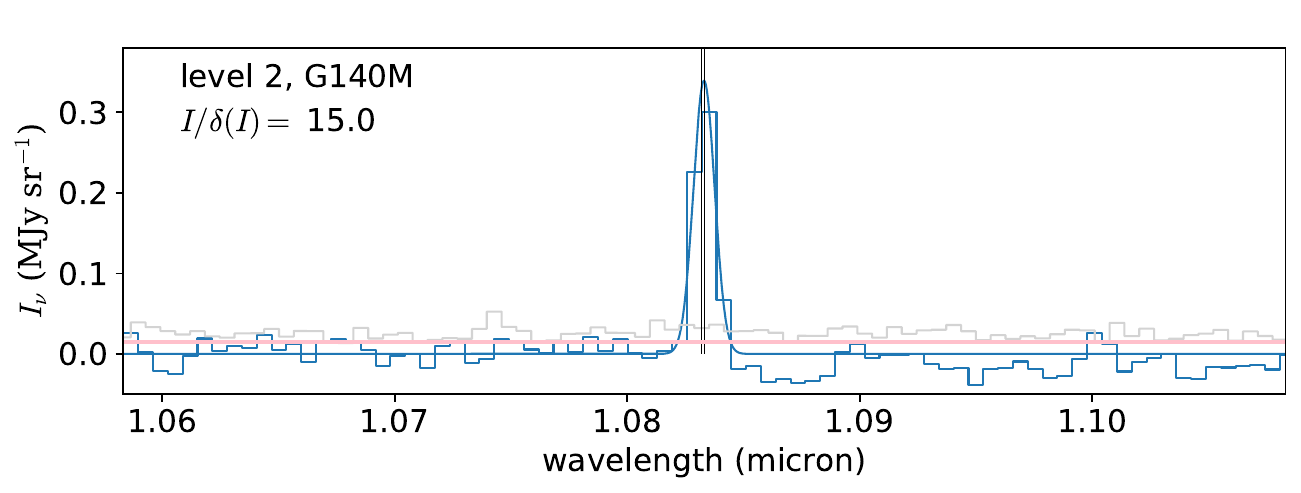}
\includegraphics[width=8.5cm]{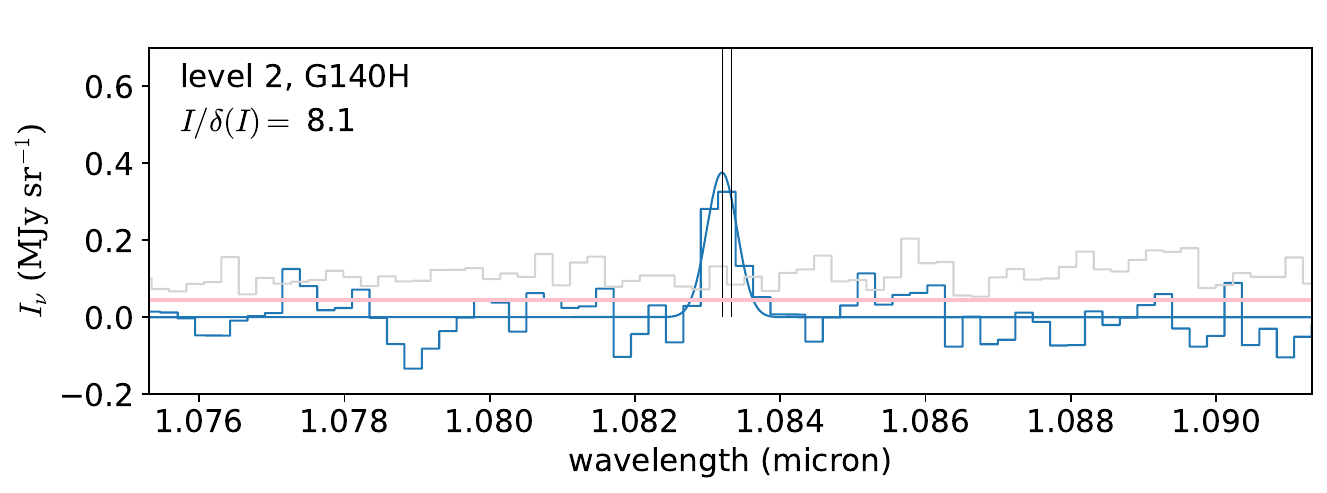}
\includegraphics[width=8.5cm]{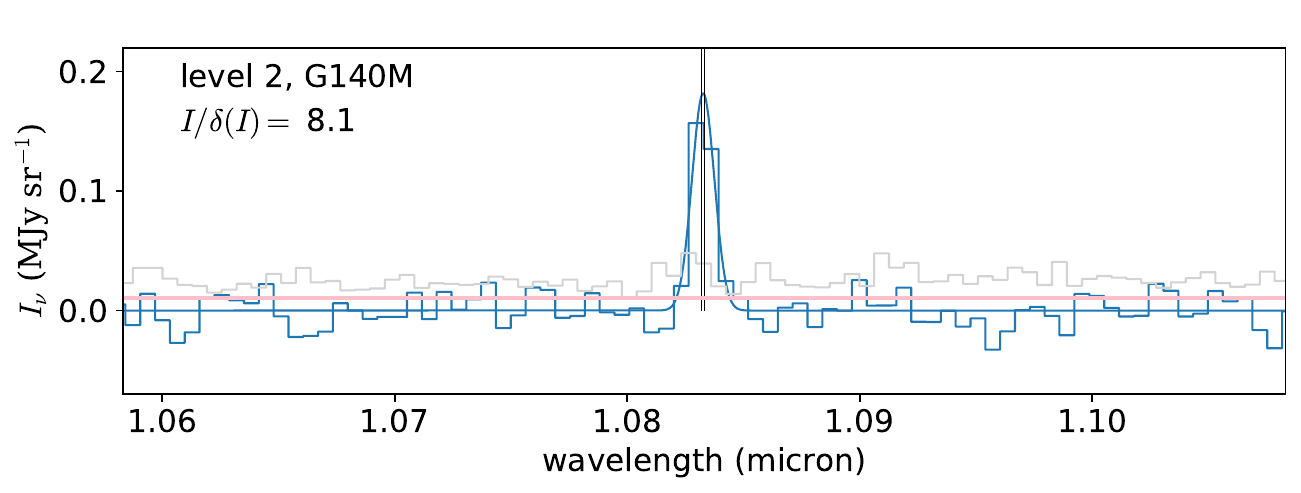}
\includegraphics[width=8.5cm]{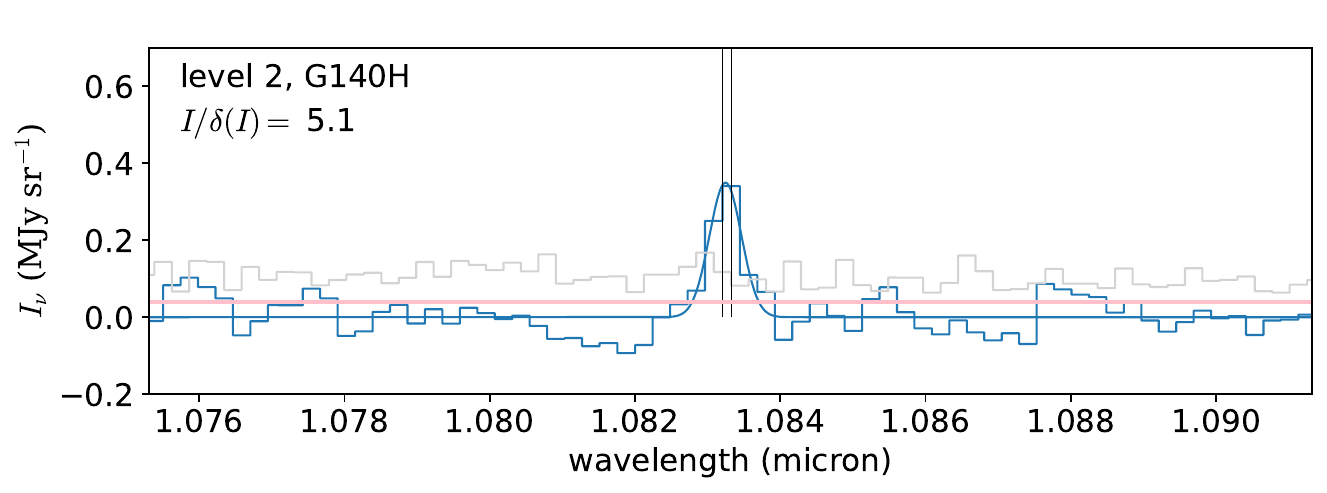}
\includegraphics[width=8.5cm]{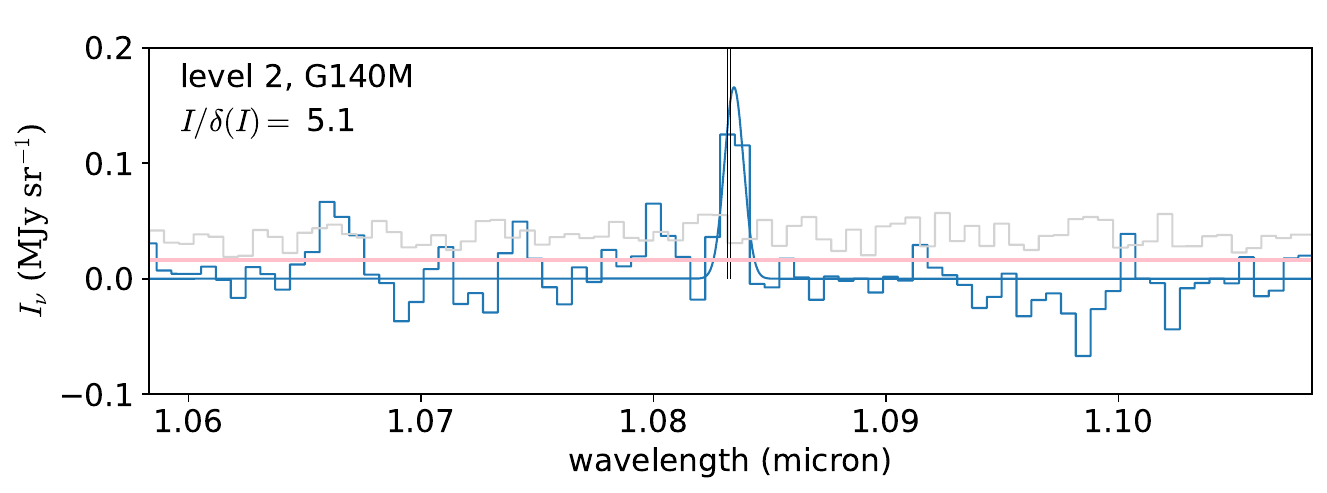}
\caption{Level 2 (exposure-level) spectra taken with the G140H (left panels) or G140M gratings (right panels) that happen to have the He I 1.0833~\micron\ emission line detected at significance levels of (top to bottom)  $I / \delta(I) = 15$, $8$, and $5$.  Spectra were taken with the 200 mas wide fixed slits.  Best fit gaussians are overplotted.   Two proxies are shown for the uncertainty in the spectrum:  the median absolute deviation in the spatial dimension (grey steps), and the standard deviation of the 1D spectrum in the dispersion direction (pink line).}
\label{fig:1DspectraSNRlevel2}  
\end{figure}

\begin{figure}
\centering
\includegraphics[width=8.5cm]{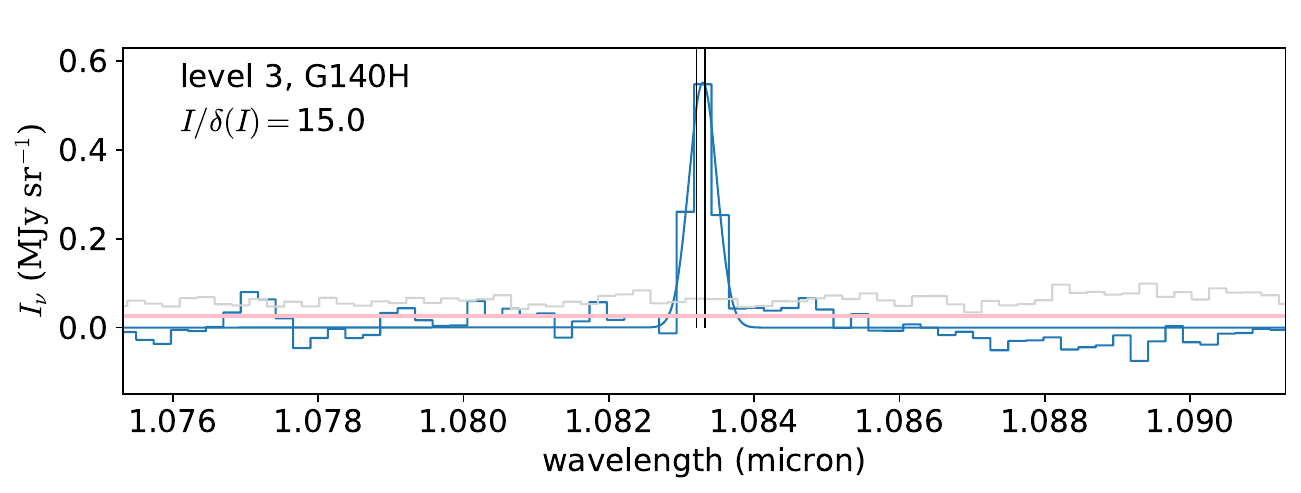}
\includegraphics[width=8.5cm]{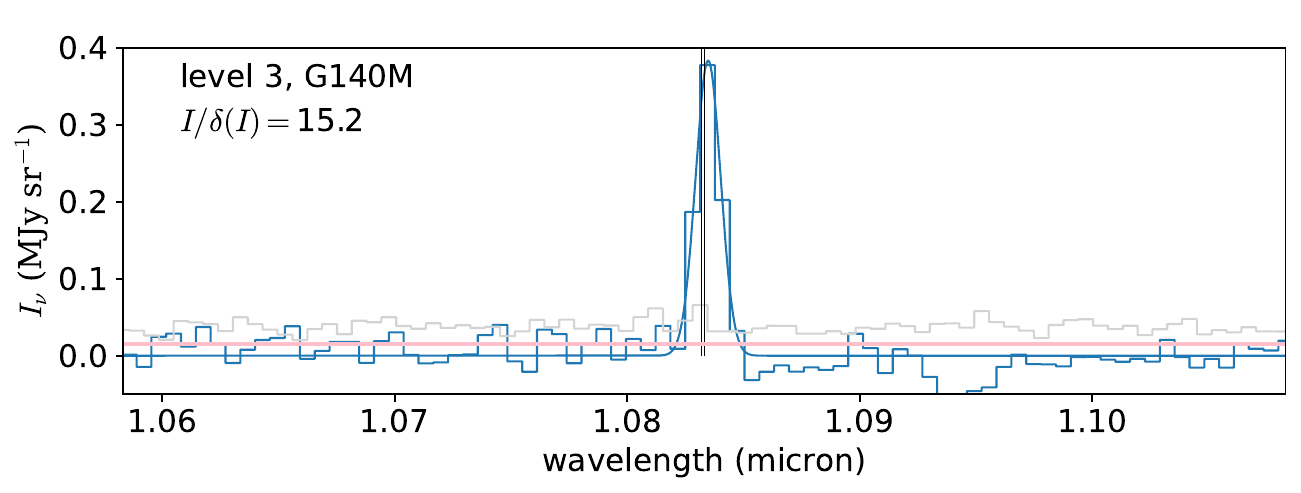}
\includegraphics[width=8.5cm]{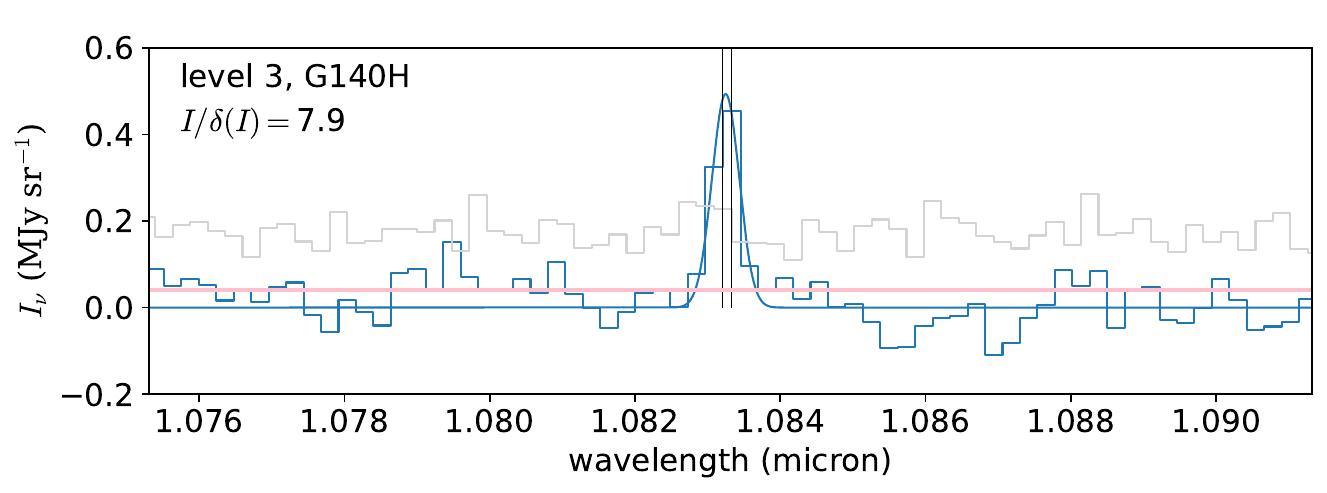}
\includegraphics[width=8.5cm]{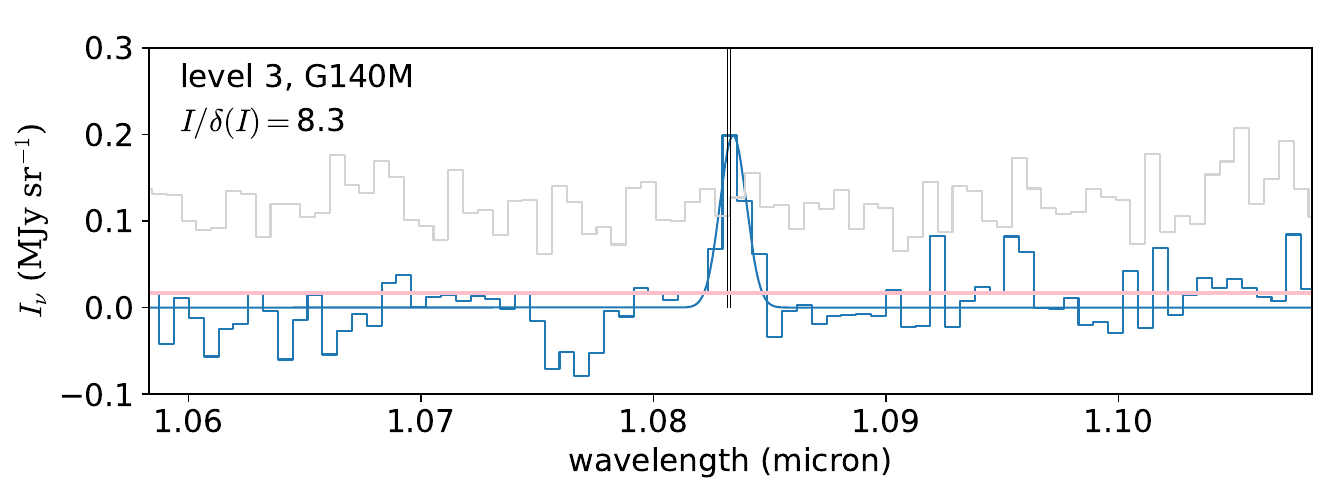}
\includegraphics[width=8.5cm]{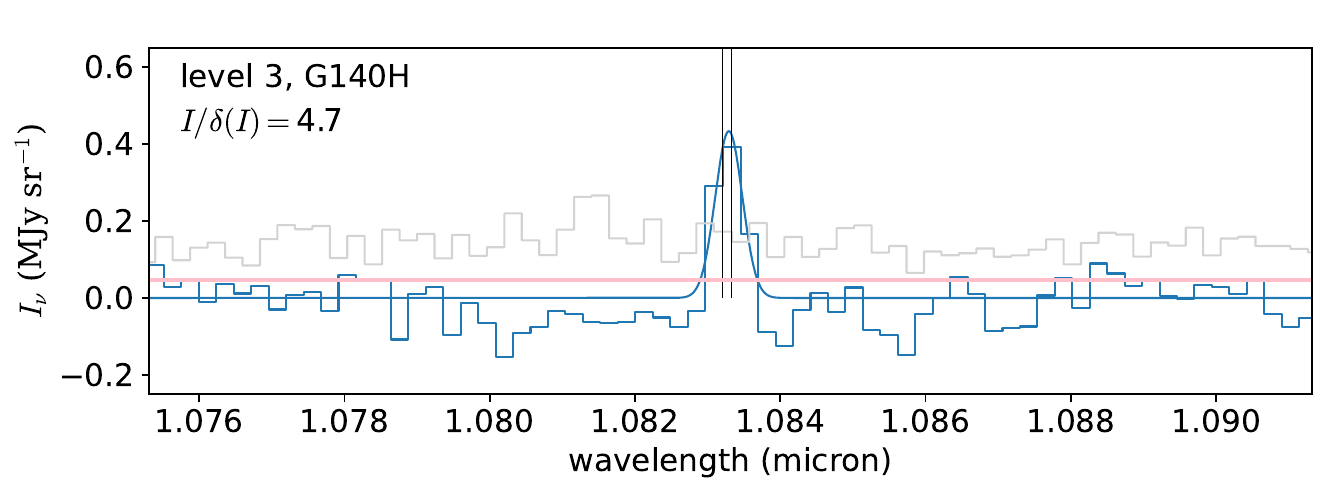}
\includegraphics[width=8.5cm]{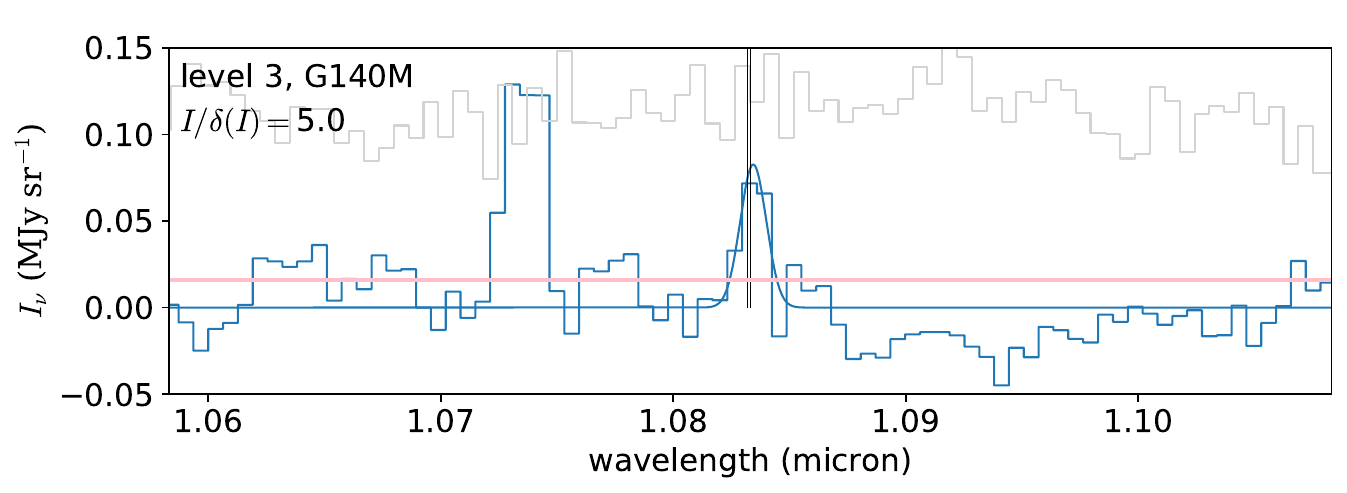}
\caption{Same as Figure~\ref{fig:1DspectraSNRlevel2}, but for the level 3 (combined) spectra.}
\label{fig:1DspectraSNRlevel3}
\end{figure}

\section{Instrumental line spread function (LSF) for NIRSpec fixed slits for uniform illumination}\label{sec:appendixLSF}
From published measurements of the line spread function (LSF) of JWST's NIRSpec instrument for point sources, we calculate the LSF for a source that uniformly illuminates the slit, as a background does.

For a point source, the LSF is simply the PSF in the dispersed direction.
The LSF is approximated by a gaussian whose FWHM is  $\delta v = c / R$ in velocity space, and
$\delta \lambda = \lambda / R$ in wavelength space.
\citet{Shajib.2025} measure the LSF for the NIRSpec fixed slits for a point source.  From the best-fit functions they publish,  we calculate the dimensionless spectral resolution for a point source $R_{pt src}$  for the He I wavelength of $\lambda = 1.0833$~\micron.  We list these values in Table~\ref{tab:R}.

\begin{figure}
\centering
\includegraphics[width=10cm]{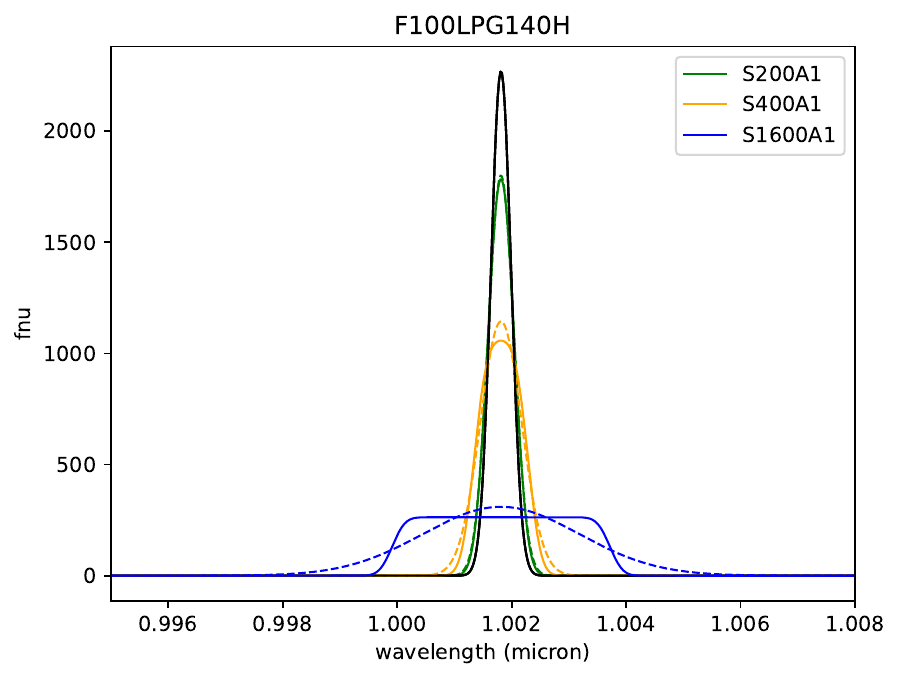}
\caption{Example of calculating the NIRSpec LSF for a source that uniformly illuminates the fixed slits.  Plotted is the LSF for a point source (black curve), and for a source that uniformly illuminates each of the fixed slits (green, yellow, and blue solid curves).  Dashed lines show the best-fit gaussian for each LSF; the single gaussian approximation is good for the two narrow fixed slits, but poor for the widest slit (S1600). This example is for the G140H grating and F100LP blocking filter.}\label{fig:LSF}
\end{figure}

For the case of a uniformly illuminated slit, the LSF is simply the convolution of the point source LSF with a tophat kernel whose width is the width of the slit in the wavelength direction.
The widths of the NIRSpec fixed slits in pixels in the wavelength direction are:
1.87 (S200A1),  1.88 (S200A2),  3.82 (S400A1), and 15.49 (S1600A1).\footnote{Values are from the JWST Science Instrument Aperture File \citep{Lallo.2017}, version PRDOPSSOC-073, accessed via pysiaf v0.28.}
We use a level 3 S2D file to determine for each pixel its width in the dispersion direction (in \micron).  We use point source LSFs from \citet{Shajib.2025}.  We then compute the LSF for a uniformly illuminated slit by performing the convolution for each wavelength (at the native pixel scale) of each NIRSpec disperser, for each fixed slit width.  To prevent under-sampling from turning the tophat kernel into a pyramid,
before convolving we oversample both the wavelength grid and the kernel by a factor of 10 compared to the native pixel scale. Figure~\ref{fig:LSF} shows examples of calculated LSFs.

We fit each calculated LSF with a single gaussian so that we may conveniently parameterize as a dimensionless R.  This single gaussian approximation works well for the S200 and S400 fixed slits, but is a poor fit to the trapezoidal LSF of the widest (S1600) fixed slit, as can be seen in Figure~\ref{fig:LSF}.   Table~\ref{tab:R} lists the calculated R and FWHM for the uniformly illuminated NIRSpec 200 mas fixed slits and the G140M and G140H gratings used in this paper.

\begin{figure}
\centering
\includegraphics[width=8.5cm]{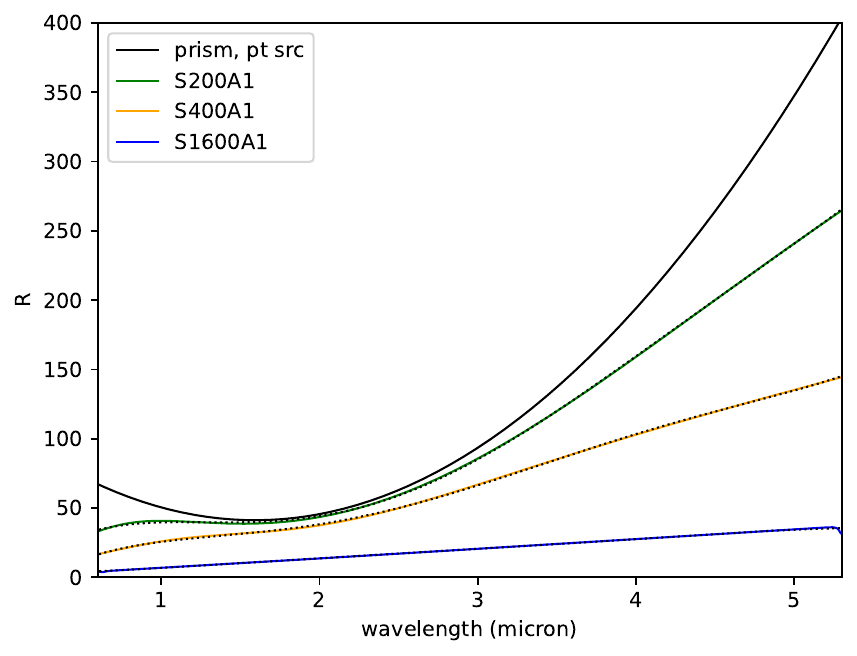}
\includegraphics[width=8.5cm]{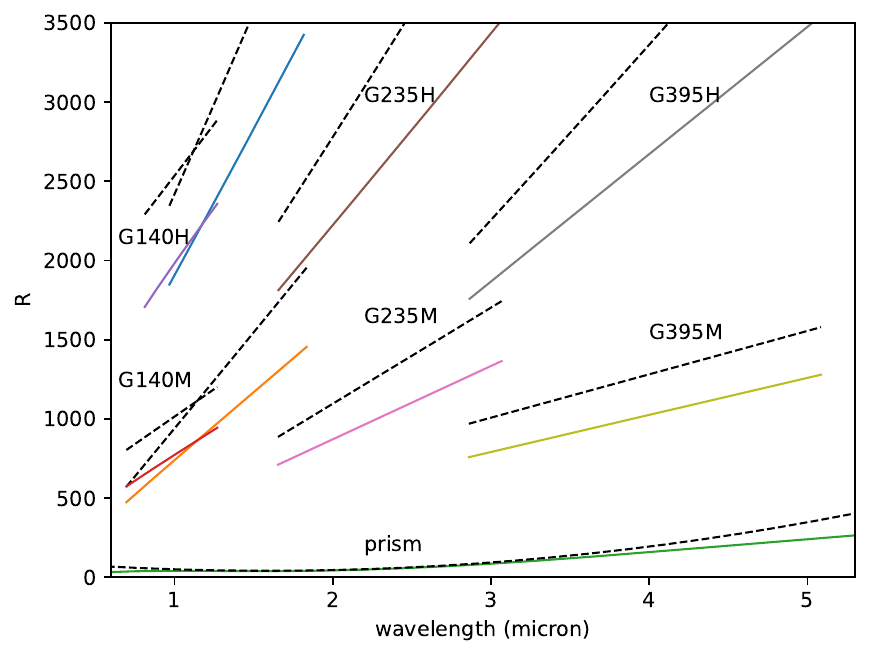}
\caption{Left panel: Spectral resolution $R$ versus wavelength for the NIRSpec prism, for a point source (top curve), and for a source that uniformly illuminates the NIRSpec fixed slits.
Right panel:  same, for a point source (black dashed lines) and for a source that uniformly illuminates the S200A1 slit (solid lines), for each of the NIRSpec dispersers.}\label{fig:Rcurve}
\end{figure}

Though this paper uses only spectra taken with the G140H and G140M dispersers, we calculate the uniformly illuminated LSF for all the NIRSpec dispersers.
See Figure~\ref{fig:Rcurve}.
To make it easy for other researchers to use these LSFs, we fit each $R(\lambda)$ curve with a polynomial:
a fifth order polynomial for the NIRSpec prism (Table~\ref{tab:Rpolyfits1}),
and a second order polynomial for the NIRSpec gratings (Table~\ref{tab:Rpolyfits2}.)
The $R(\lambda)$ we calculate for the 200 mas wide fixed slit should be appropriate for the NIRSpec microshutters, since they have the same width.

\begin{deluxetable}{lccccc}
\tablecaption{Spectral resolution for NIRSpec at 1.0833~\micron\label{tab:R}}
\tablehead{
\colhead{Filter} &
\colhead{Grating} &
\colhead{$R_{pt src}$} &
\colhead{$R_{uniform}$} &
\colhead{$FWHM_{pt src}$  (\kms)} &
\colhead{$FWHM_{uniform}$ (\kms)}
}
\startdata
F070LP & G140H & 2642 & 2100 & 113 & 143\\ 
F100LP & G140H & 2605 & 2060 & 115 & 146\\ 
F070LP & G140M & 1071 & 826  & 280 & 363\\ 
F100LP & G140M & 1039 & 811  & 289 & 370\\ 
\enddata
\tablecomments{Spectral resolution as dimensionless $R$, and as a FWHM in velocity space,
calculated for the wavelength of 1.0833~\micron, for the NIRSpec 200 mas fixed slits, for two cases:
  a point source (from \citealt{Shajib.2025}), and uniform illumination (from Appendix~\ref{sec:appendixLSF}).}
\end{deluxetable}

\begin{deluxetable}{lccccccccc}
\tabletypesize{\scriptsize}
\tablecaption{Polynomial fits to NIRSpec's spectral resolution for uniform illumination, for the prism\label{tab:Rpolyfits1}}
\tablehead{
\colhead{Filter} &
\colhead{disperser} &
\colhead{slitname} &
\colhead{order} &
\colhead{c0} &
\colhead{c1} &
\colhead{c2} &
\colhead{c3} &
\colhead{c4} &
\colhead{c5}}
\startdata
 --     &  prism & S200A1  & 5 & -8.84463374   & 130.23476077  & -125.66897502 & 51.88026451 & -8.49676159 & 0.50090293\\
 --     &  prism & S400A1  & 5 & -22.96738706  & 108.13587648  & -91.01351902  & 37.59929141 & -6.56990109 & 0.41556722\\
 --     &  prism & S1600A1 & 5 & 2.02447098    & 1.21353671    & 5.50963087    & -2.42974433 & 0.49096522  & -0.03667465\\
\enddata
\tablecomments{Polynomial fits to computed curves of spectral resolution for uniformly illuminated NIRSpec fixed slits, 
  $R(\lambda) =  c0 + c1x + c2x^2 + c3x^3 + c4x^4 + c5x^5$, where x is wavelength in \micron. Order is the order of the polynomial.}
\end{deluxetable}

\begin{deluxetable}{lccccccc}
\tabletypesize{\scriptsize}
\tablecaption{Polynomial fits to NIRSpec's spectral resolution for uniform illumination, for the gratings\label{tab:Rpolyfits2}}
\tablehead{
\colhead{Filter} &
\colhead{disperser} &
\colhead{slitname} &
\colhead{order} &
\colhead{c0} &
\colhead{c1} &
\colhead{c2}}
\startdata
 F100LP &  G140H & S200A1  & 2 & 111.93311684  & 1758.73370242 & 35.21053235\\
 F100LP &  G140H & S400A1  & 2 & 19.57172657   & 1155.76855775 & 6.13012243\\
 F100LP &  G140H & S1600A1 & 2 & -0.93352317   & 312.96219031  & -0.48109856\\
 F100LP &  G140M & S200A1  & 2 & -138.70824899 & 889.82263714  & -12.12140729\\
 F100LP &  G140M & S400A1  & 2 & -27.19059316  & 490.11565067  & -4.90167502\\
 F100LP &  G140M & S1600A1 & 2 & -0.66575965   & 121.7784199   & -0.27191219\\
 F070LP &  G140M & S200A1  & 2 & 93.35846682   & 715.57225091  & -36.62291343\\
 F070LP &  G140M & S400A1  & 2 & 9.11808216    & 464.87640324  & -10.52979768\\
 F070LP &  G140M & S1600A1 & 2 & -1.89090696   & 124.89872615  & -2.08962618\\
 F070LP &  G140H & S200A1  & 2 & 373.67387436  & 1785.08130901 & -176.78040094\\
 F070LP &  G140H & S400A1  & 2 & 25.61699616   & 1259.87116199 & -66.40665676\\
 F070LP &  G140H & S1600A1 & 2 & -0.31016457   & 322.22312353  & -1.25535672\\
 F170LP &  G235H & S200A1  & 2 & -118.14087797 & 1142.10636419 & 13.96292928\\
 F170LP &  G235H & S400A1  & 2 & -24.77878853  & 707.96266443  & 1.77946606\\
 F170LP &  G235H & S1600A1 & 2 & -1.4993786    & 186.59261563  & -0.18800551\\
 F170LP &  G235M & S200A1  & 2 & -53.74441013  & 461.65864235  & 0.3000743\\
 F170LP &  G235M & S400A1  & 2 & -10.19419108  & 277.57514132  & -0.09291406\\
 F170LP &  G235M & S1600A1 & 2 & -1.64489179   & 73.39176608   & -0.25930059\\
 F290LP &  G395H & S200A1  & 2 & -537.02702138 & 800.74334325  & 0.32981391\\
 F290LP &  G395H & S400A1  & 2 & -128.43888328 & 456.83099287  & -1.88822041\\
 F290LP &  G395H & S1600A1 & 2 & -1.51264255   & 110.85059056  & -0.02927819\\
 F290LP &  G395M & S200A1  & 2 & 83.19818524   & 237.16845995  & -0.39640376\\
 F290LP &  G395M & S400A1  & 2 & 13.055894     & 159.03822564  & -0.20785423\\
 F290LP &  G395M & S1600A1 & 2 & -1.8061062    & 43.78160899   & -0.1179814\\
\enddata
\tablecomments{Polynomial fits to computed curves of spectral resolution for uniformly illuminated NIRSpec fixed slits, 
  $R(\lambda) =  c0 + c1x + c2x^2$, where x is wavelength in \micron.  Order is the order of the polynomial.}
\end{deluxetable}

\section{When does Earth cross the middle of the helium focusing cone?}\label{sec:appendixconecenter}
The literature gives the date when the Earth passes through the helium focusing cone as  ``early December'' \citep{Mobius.2004, Galeazzi.2014}.  We calculate that date to the greater precision needed for the analysis in this paper. We  take the Ecliptic coordinates for the wind direction as
$74.5$\degr\ ($\pm  0.5$\degr ), $-5.7$\degr\ ($\pm  0.5$\degr )
(longitude and latitude), which were reported by \citet{Mobius.2004} as a concordance of several measurements. Assuming the cone is directly downwind, it should be oriented along 254.5\degr\ ($\pm 0.5$\degr), $+5.7$\degr\ ($\pm 0.5$\degr) Ecliptic longitude and latitude.  This position is consistent within uncertainties with the cone axis measured from eROSITA observations \citep{Dennerl.2026}. 
Converting this downwind direction to equatorial coordinates gives the right ascension (RA) of the sun when the Earth is in the center of the cone: 16h55m28s. Using the observing tool JskyCalc \citep{skycalc}, we calculate that for an observer on the Earth, the sun has this RA at time UT 2025 Dec.\ 07 00:11:00. The $\pm 0.5$\degr\ uncertainty in longitude reported by \citet{Mobius.2004} propagates to a $\sim 12$ hr uncertainty in time.  We ignore the $\sim 0.01$ AU distance between the Earth and JWST.

\end{document}